\documentclass[aps,prl,reprint,superscriptaddress]{revtex4-2}
\usepackage{amsmath}
\usepackage{amssymb}
\usepackage{mathtools}
\usepackage{graphicx}
\usepackage{dcolumn}
\usepackage{bm}
\usepackage[dvipsnames]{xcolor}
\usepackage[hidelinks,colorlinks=true,allcolors=NavyBlue]{hyperref}
\usepackage[capitalise]{cleveref}
\usepackage[siunitx]{circuitikz}
\usepackage{physics}
\usepackage{float}
\usepackage{adjustbox}
\usepackage[caption=false]{subfig}
\renewcommand{\a}{\hat{a}}
\newcommand{\D}{\hat{D}}
\newcommand{\x}{\hat{x}}
\newcommand{\p}{\hat{p}}
\renewcommand{\H}{\hat{H}}

\newcommand{\U}{\hat{U}}

\newcommand{\rhoh}{\hat{\rho}}

\usepackage{pdfpages} 
\usepackage{pgffor} 

\makeatletter
\AtBeginDocument{\let\LS@rot\@undefined}
\makeatother

\def\supplementfilename{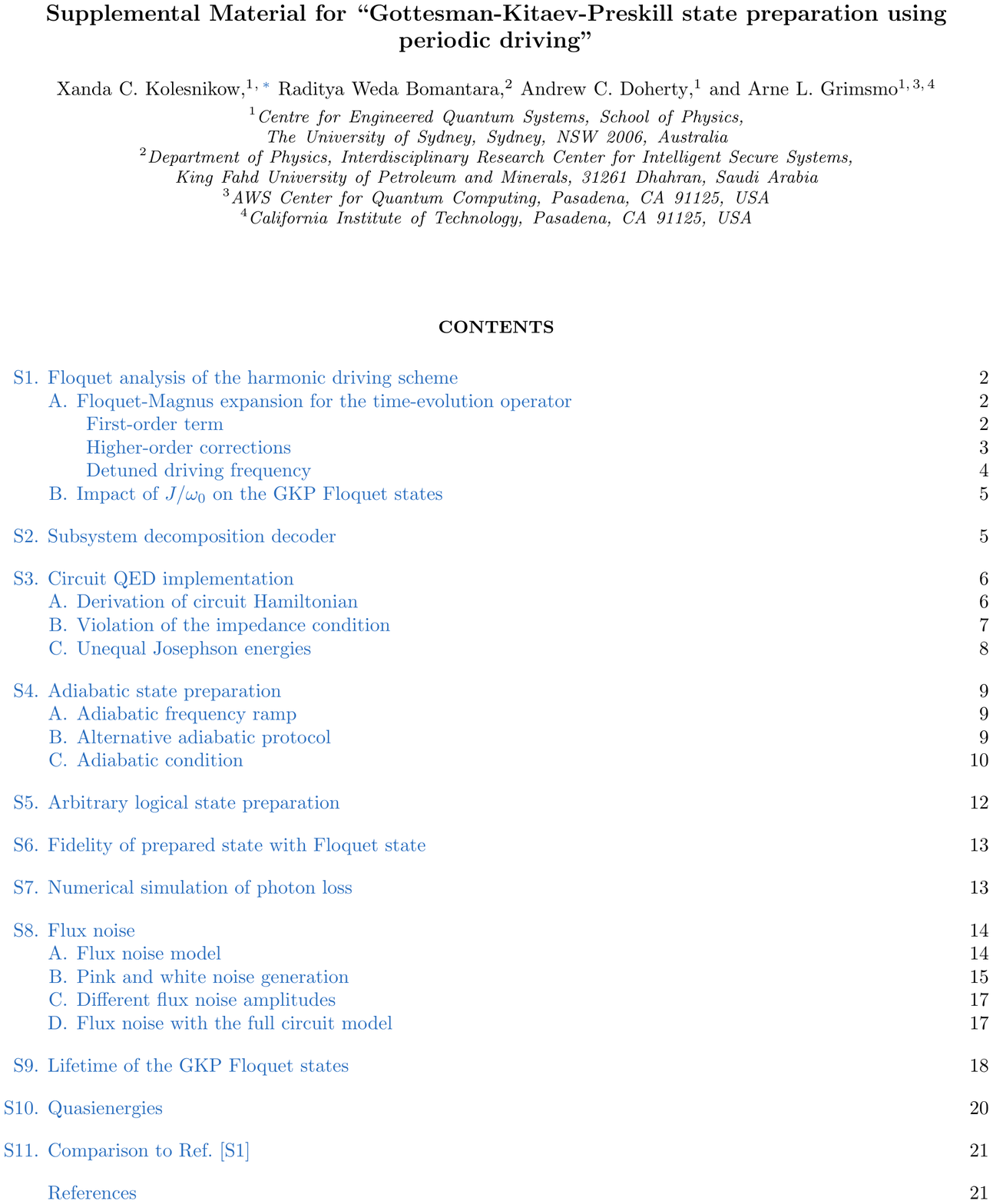}

\pdfximage{\supplementfilename}
\def\numbersupplementpages{\the\pdflastximagepages}

\newif\ifarXiv
\arXivtrue 

\begin{document}
	
	\title{Gottesman-Kitaev-Preskill state preparation using periodic driving}
	
	\author{Xanda C. Kolesnikow}
	\email{xkol5366@uni.sydney.edu.au}
	\affiliation{Centre for Engineered Quantum Systems, School of Physics, The University of Sydney, Sydney, NSW 2006, Australia}
	\author{Raditya W. Bomantara}
	\affiliation{Department of Physics, Interdisciplinary Research Center for Intelligent Secure Systems, King Fahd University of Petroleum and Minerals, 31261 Dhahran, Saudi Arabia}
	\author{Andrew C. Doherty}
	\affiliation{Centre for Engineered Quantum Systems, School of Physics, The University of Sydney, Sydney, NSW 2006, Australia}
	\author{Arne L. Grimsmo}
	\affiliation{Centre for Engineered Quantum Systems, School of Physics, The University of Sydney, Sydney, NSW 2006, Australia}
	\affiliation{AWS Center for Quantum Computing, Pasadena, CA 91125, USA}
        \affiliation{California Institute of Technology, Pasadena, CA 91125, USA}
	
	\date{\today}
	
	\begin{abstract}
		The Gottesman-Kitaev-Preskill (GKP) code may be used to overcome noise in continuous variable quantum systems. However, preparing GKP states remains experimentally challenging. We propose a method for preparing GKP states by engineering a time-periodic Hamiltonian whose Floquet states are GKP states. This Hamiltonian may be realized in a superconducting circuit comprising a SQUID shunted by a superinductor and a capacitor, with a characteristic impedance twice the resistance quantum. The GKP Floquet states can be prepared by adiabatically tuning the frequency of the external magnetic flux drive. We predict that highly squeezed $>11.9$~dB ($10.8$~dB) GKP magic states can be prepared on a microsecond timescale, given a quality factor of $10^6$ ($10^5$) and flux noise at typical rates.
	\end{abstract}
	
	\maketitle
	
	The Gottesman-Kitaev-Preskill (GKP) code encodes discrete variable quantum information into continuous variable quantum systems, and may be used to protect the encoded information from noise~\cite{Gottesman2001,Albert2018,Terhal2020,Grimsmo2021}. GKP states have recently been prepared and stabilized experimentally in trapped ion~\cite{Fluehmann2019,Neeve2022} and circuit quantum electrodynamics (QED) architectures~\cite{CampagneIbarcq2020,Eickbusch2022a,Sivak2023}, using an ancillary qubit to perform phase estimation of the GKP code stabilizers~\cite{Terhal2016,Weigand2020,Royer2020,Grimsmo2021}. However, preparing high-quality GKP states remains a central challenge for practical quantum error correction with GKP codes.

	An alternative approach has been proposed theoretically, where GKP states arise naturally as eigenstates of superconducting circuits with a carefully engineered phase-charge symmetry~\cite{Douccot2012,Le2019,Rymarz2021}. This passive approach is appealing, but the proposals rely on exotic circuit elements that may be hard to realize in practice. In Ref.~\cite{Conrad2021}, it was recently shown that circuit complexity can be traded for time-dependent control, by demonstrating how GKP states can arise as Floquet states of a periodically displaced harmonic oscillator (HO). Here, the price to pay is instead precise, high-frequency control, as the scheme relies on periodic, near-instantaneous displacements of the HO state [see the Supplemental Material (SM)~\cite{SM} for a detailed comparison to this work].
	
	Building on the idea of trading hardware for control complexity, we propose an ancilla-free method based on modulating a nonlinear element. This scheme may be realized, for example, with a SQUID shunted by a superinductor and a capacitor. We show that with this nonlinear control mechanism, a small number of harmonics of the circuit frequency suffices for the control signal. Numerical simulations indicate that GKP states with large squeezing levels $>11.9$~dB ($10.8$~dB) can be realized even in the presence of flux noise at realistic rates and photon loss assuming a quality factor of $10^6$ ($10^5$).
	
	In this Letter we focus on the single-mode square GKP code, which has commuting stabilizer generators $\D(\sqrt{2\pi})$ and $\D(\sqrt{2\pi}i)$, where $\D(\alpha) = e^{\alpha \a^\dag - \alpha^* \a}$ is the displacement operator for a bosonic mode with annihilation operator $\a$, satisfying $[\a, \a^\dag]=1$. Equivalently, the codespace may be defined as the twofold degenerate ground space of the GKP Hamiltonian~\cite{Gottesman2001}
	\begin{equation}
			\H_\text{GKP}/\hbar = -J\big(\cos(2\sqrt{\pi}\x) + \cos(2\sqrt{\pi}\p) \big), \label{eqn:Hgkp}
	\end{equation}
	where $\hbar J$ defines the energy scale and $\x = (\a^\dag + \a)/\sqrt{2}$ and $\p = i(\a^\dag - \a)/\sqrt{2}$ are the dimensionless position and momentum quadratures. Note that the GKP Hamiltonian is a special case of the Harper model~\cite{Harper1955,Aubry1980}, which is of importance in studies of quantum chaos and topological insulators~\cite{Hofstadter1976,Thouless1982,Leboeuf1990,Aidelsburger2014}.
 
    As the eigenstates of the GKP Hamiltonian are not normalizable, they cannot be prepared exactly. Instead one must resort to states that approximate ideal GKP states~\cite{Gottesman2001}. To characterize how close a state is to an ideal GKP state, we use the GKP squeezing parameter~\cite{Duivenvoorden2017} 
	\begin{equation}
		\Delta = \sqrt{-\frac{1}{2\pi}\log(|\langle\D(\sqrt{2\pi})\rangle|^2)}, \label{eqn:SqueezingParameter}
	\end{equation}
	which we convert to decibels (dB) via $\mathcal{S} = -10\log_{10}(\Delta^2)$, and refer to as squeezing for short. Alternatively, the squeezing may be measured with respect to the orthogonal stabilizer generator $\D(i\sqrt{2\pi})$, but for symmetrically squeezed states these two values are equal. Note that $\Delta$ approaches zero ($\mathcal S \to \infty$) as $|\langle\D(\sqrt{2\pi})\rangle|$ approaches one and the state approaches the GKP codespace. Numerical studies suggest that squeezing in the range $10$-$12$~dB is required for scalable error correction~\cite{Menicucci2014,Vuillot2019,Noh2020,Noh2022}.
	
	Whilst the squeezing quantifies how close an arbitrary state is to the GKP codespace, it does not give any information about the \textit{logical} information contained in the state. To quantify this, we use the subsystem decomposition introduced in Ref.~\cite{Shaw2024}. Under this decomposition, a partial trace operation amounts to an ideal decoder $\mathcal{D}$ that takes density matrices in the bosonic Hilbert space to logical $2\times 2$ density matrices. This can be used to introduce a \textit{logical} fidelity metric, by computing the fidelity of the decoded state to a target logical state.
	
	In the following, we introduce an approximate form of the GKP Hamiltonian, arising as an effective stroboscopic Hamiltonian in a periodically driven system. At its core, our scheme substitutes the requirement of periodicity in two conjugate variables [cf.~\cref{eqn:Hgkp}], with periodicity in one variable \textit{and time}. Our approach utilizes similar techniques to Floquet engineering topological phases via driving otherwise trivial and simple systems~\cite{Oka2009,Lindner2011}, and extends these tools to the domain of quantum error correction.

	\textit{The kicked HO.---}Consider the time-periodic Hamiltonian 
	\begin{equation}
		\H(t)/\hbar = \omega_0 \a^\dag \a - J f(t) \cos(2\sqrt{\pi} \x), \label{eqn:H}
	\end{equation} 
	with the periodic driving function $f(t) = \frac{T}{2}\sum_{n=0}^\infty \delta(t - nT/4)$, where $\delta(t)$ denotes the Dirac delta function, $\omega_0$ is the HO frequency, and $T$ is the driving period~\footnote{Note that $f(t+T/4) = f(t)$ so that the minimal period for $\H(t)$ is $T/4$. However, the following results are most easily seen by defining $T$ to be the period for $\H(t)$ and considering time evolution over this period.}. This ``kicked HO'' model, which represents a freely evolving HO interspersed with four delta-function kicks within one driving period, was previously studied for exploring many-body topological phases~\cite{Liang2018}. The one-period time-evolution operator (referred to as the Floquet operator) for this model can be written as $\U_T = \left(e^{-i\omega_0 \a^\dag \a T/4} e^{-iJ\cos(2\sqrt{\pi}\x)T/2}\right)^4$.
	
	If the HO frequency is made equal to the driving frequency ($\omega_0 = 2\pi/T$), then via the formula $e^{i\theta \a^\dag \a} \x e^{-i\theta \a^\dag \a} = \x\cos\theta + \p\sin\theta$, we may rotate half of the $\cos(2\sqrt{\pi}\x)$ terms to $\cos(2\sqrt{\pi}\p)$ terms. In this special case, the one-period time evolution becomes equivalent to a kicked Harper model~\cite{Zaslavskii1986,Wang2013}. Furthermore, since $\cos(2\sqrt{\pi}\x)$ commutes with $\cos(2\sqrt{\pi}\p)$, the Floquet operator may be written as $\U_T = e^{-iT\H_\text{GKP}/\hbar}$. That is, time evolution under the time-dependent Hamiltonian for a single period is equivalent to evolving under the static GKP Hamiltonian in \cref{eqn:Hgkp} for the same amount of time. This means that the eigenstates of the Floquet operator, referred to as Floquet states, are the ideal GKP states.
	
\begin{figure}[t]
        \subfloat{\label{fig:tvsf}}
		\subfloat{\label{fig:NvsS}}
		\subfloat{\label{fig:NvsFlog}}
		\subfloat{\label{fig:WignerPlus}}
		\subfloat{\label{fig:WignerMinus}}
		\centering
		\includegraphics[width=\linewidth]{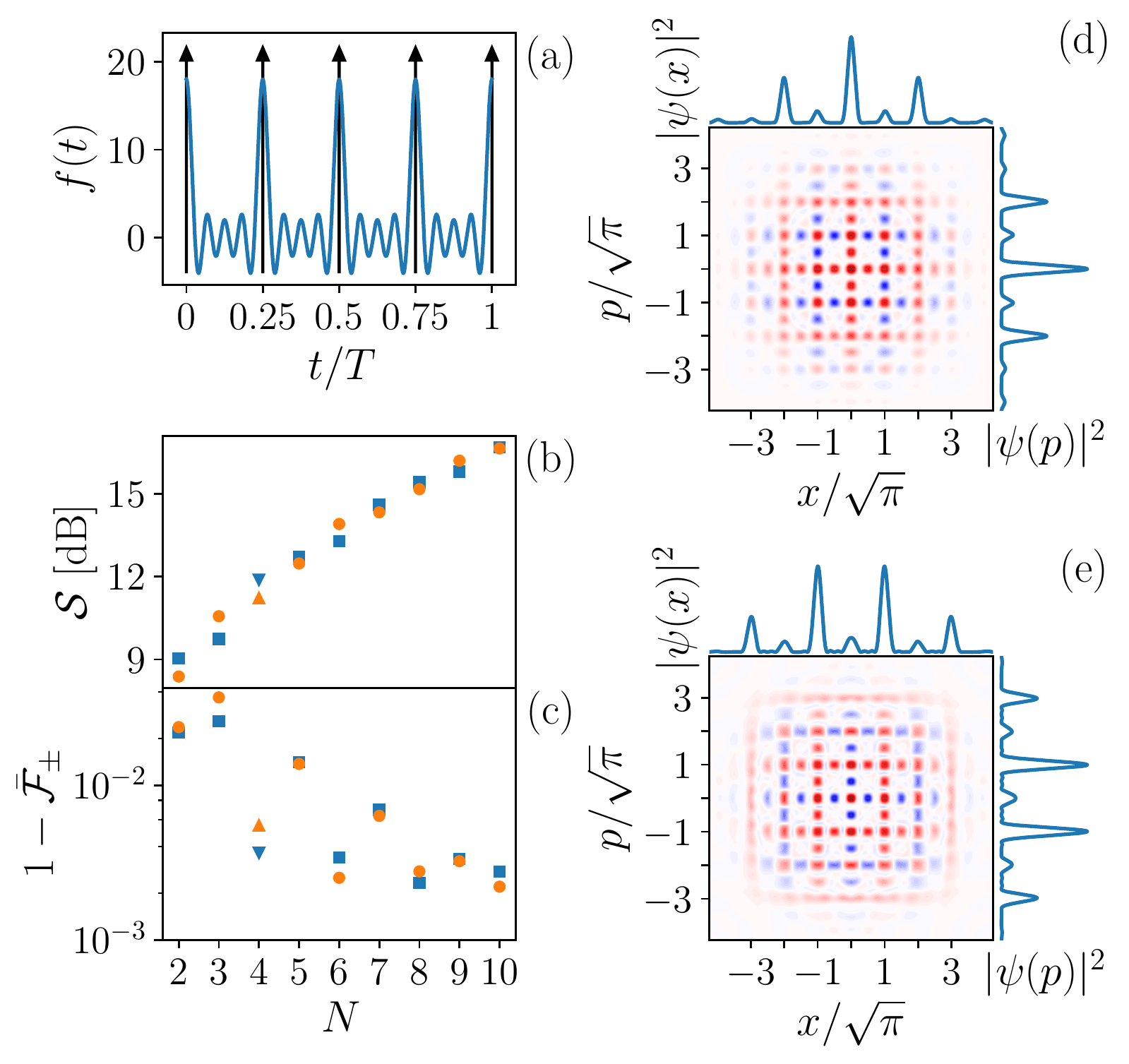}
		\caption{The harmonic driving scheme. (a) The harmonic driving function with $N=4$ harmonics (blue), and in the $N \to \infty$ limit (black). Arrows represent Dirac delta functions. Squeezing (b) and logical fidelity (c) of the $\ket*{\psi^{(N)}_\pm}$ Floquet states as a function of the number of harmonics $N$. Blue squares (orange circles) denote values for the $\ket*{\psi_+^{(N)}}$ ($\ket*{\psi_-^{(N)}}$) state. (d)  [(e)] Wigner function and marginal probability distributions for $\ket*{\psi^{(4)}_+}$ ($\ket*{\psi^{(4)}_-}$). Red (blue) represents positive (negative) values of the Wigner function. The blue downward (orange upward) pointing triangles in (b) and (c) correspond to the parameter values for the Floquet state in (d) [(e)]. All results are for $J/\omega_0 = 2.5 \times 10^{-3}$.}
		\label{fig:hdscheme}
	\end{figure}

	\textit{Harmonic driving scheme.---}Exactly realizing the delta-function drive is impractical. In the following we introduce a new periodically driven model with a harmonic driving scheme, motivated by the kicked HO model. We start by truncating the Fourier series decomposition of the delta-function drive at a finite number of harmonics $N$ [\cref{fig:tvsf}], to obtain
	\begin{equation}
		f(t) = 2 + 4\sum_{n=1}^N \cos(4n \omega_0 t). \label{eqn:f}
	\end{equation}
	It may then be shown that for fixed $N$, the Floquet operator takes the form $\U_T = e^{-iT\H_F/\hbar}$, where 
	\begin{equation}
		\H_F/\hbar = \H_\text{GKP}^{(N)}/\hbar + O\bigl(J^2/\omega_0\bigr), \label{eqn:HF}
	\end{equation}
	and $\H_\text{GKP}^{(N)}$ is the GKP Hamiltonian [\cref{eqn:Hgkp}] truncated at the $\pm 4N$ diagonals in the Fock basis such that $\H_\text{GKP} = \lim_{N\to\infty} \H_\text{GKP}^{(N)}$. Thus, to zeroth order in $J^2/\omega_0$, the effective static Hamiltonian converges to the GKP Hamiltonian in the $N \to \infty $ limit. The details of this calculation are given in the SM~\cite{SM}.

	The truncation in Fock space leads to two nearly degenerate approximate GKP ground states whose squeezing increases with the number of harmonics. Furthermore, the truncated GKP Hamiltonian has Fourier transform symmetry for all $N$, that is $[\H^{(N)}_\text{GKP}, e^{i\pi \a^\dag \a/2}] = 0$, meaning that its eigenstates are invariant under a Fourier transform. Since the Fourier transform operator acts as the Hadamard gate on a square GKP code~\cite{Gottesman2001}, the two nearly degenerate ground states are approximate GKP magic states~\cite{Bravyi2005}. Furthermore, this symmetry ensures that these states are equally squeezed in both quadratures, justifying our choice of metric in \cref{eqn:SqueezingParameter}.  
	
	Consistent with the correction term in \cref{eqn:HF}, for small $J/\omega_0$, the Floquet states inherit the properties of the eigenstates of $\H_\text{GKP}^{(N)}$ (in the SM~\cite{SM} we quantify this). To quantify the logical information of the Floquet states, we compute their logical fidelity with the Hadamard eigenstates, $\bar{\mathcal{F}}_\pm = \max_i\bra{H_\pm} \mathcal{D}\bigl[\ketbra*{\psi_i^{(N)}}{\psi_i^{(N)}}\bigr]\ket{H_\pm}$,
	where $\ket{H_\pm}$ are the $\pm 1$ eigenstates of the $2 \times 2$ Hadamard gate, $\ket*{\psi_i^{(N)}}$ are the Floquet states for an $N$-harmonic driving scheme, and  we denote by $\ket*{\psi^{(N)}_\pm}$ the Floquet states that achieve this maximum fidelity (the SM~\cite{SM} contains details on how we numerically implement the decoder $\mathcal D$). Given that the $\ket{H_\pm}$ states are magic states, we may also view this metric as quantifying the distillability of the Floquet states~\cite{Bravyi2005}.  
 
    Numerical results for $J/\omega_0 = 2.5 \times 10^{-3}$ are shown in \cref{fig:NvsS,fig:NvsFlog,fig:WignerPlus,fig:WignerMinus}. Here we find that the squeezing of the $\ket*{\psi_\pm^{(N)}}$ Floquet states increases monotonically with $N$ [\cref{fig:NvsS}], and that the logical fidelity generally increases with $N$, albeit not monotonically~[\cref{fig:NvsFlog}]. In particular, the $\ket*{\psi^{(4)}_+}$ ($\ket*{\psi_-^{(4)}}$) state, whose Wigner function and marginal probability distributions are shown in \cref{fig:WignerPlus} [\cref{fig:WignerMinus}], has logical infidelity $3.7\times10^{-3}$ ($5.5\times 10^{-3}$) and squeezing $11.9$~dB ($11.2$~dB). This suggests that $N=4$ harmonics is sufficient to generate high-quality GKP states. In the next section we discuss how these parameter may be attained in a circuit QED realization.
	
	\textit{Circuit QED implementation.---}Whilst the driving scheme could be implemented in a variety of quantum computing architectures -- in particular cold atoms may be an intriguing platform~\cite{Liang2018} -- here we focus on a proposal using superconducting circuits~\cite{Blais2021}. We reserve the derivation of the circuit Hamiltonian for the SM~\cite{SM} and outline the main physical requirements here. 
	
\begin{figure}[t]
		\subfloat{\label{fig:CircuitDiagram}}
		\subfloat{\label{fig:AdiabaticRamps}}
		\centering
            \includegraphics[width=\linewidth]{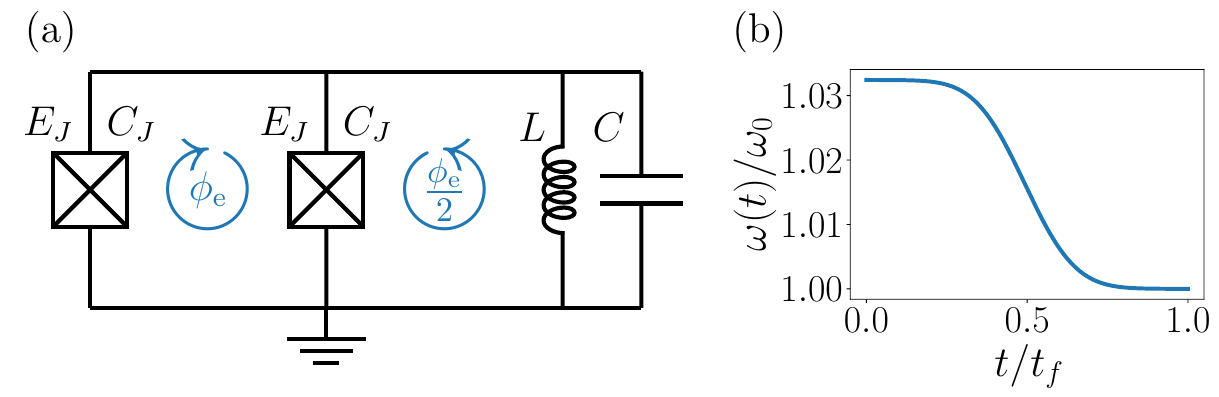}%
		\caption{(a) Circuit diagram for the harmonic driving scheme. The circuit comprises a symmetric SQUID loop with Josephson energies $E_J$ and capacitances $C_J$ shunted by a linear inductor with inductance $L$ and capacitor with capacitance $C$. The two loops in the circuit are threaded by external fluxes $\phi_{\text{e}}$ and $\phi_{\text{e}}/2$ in opposite directions (b) Frequency ramp for the external flux drive to prepare GKP states with a preparation time $t_f$. The driving is switched off at $t=t_f$ once the state is prepared (see discussion and outlook section).}
	\end{figure}

	\Cref{eqn:H} can be realized as a symmetric SQUID loop shunted by a linear inductor and a capacitor [\cref{fig:CircuitDiagram}] with circuit frequency $1/\sqrt{L C_\Sigma} = \omega_0$ and circuit impedance $\sqrt{L/C_\Sigma} = 2R_Q$, where $C_\Sigma = 2C_J + C$ is the total capacitance in the circuit, and $R_Q = h/(2e)^2$ is the resistance quantum. We note that the same impedance condition was necessary to realize the schemes in Refs.~\cite{Conrad2021,Rymarz2021}. Although an impedance surpassing the resistance quantum is difficult to realize with a conventional LC oscillator, impedances larger than $2 R_Q$ have been obtained using chains of Josephson junctions~\cite{Masluk2012,Kuzmin2019,Kuzmin2021,Pechenezhskiy2020,Smith2022}, thin-film disordered superconductor nanowires~\cite{Hazard2019}, and suspended aluminium coils~\cite{Peruzzo2020}. As for the robustness to deviations from $2R_Q$, we observe a decrease of no more than $0.7$~dB ($4 \times 10^{-3}$) in the squeezing (logical fidelity) of the GKP Floquet states for variations of $\pm 0.1 R_Q$~(see the SM~\cite{SM} for more details).
	
	In order to implement the driving function, the external flux threading the SQUID loop is varied as 	
	\begin{equation}
		\phi_{\text{e}}(t)/\phi_0 = \pi - \epsilon f(t), \label{eqn:phix}
	\end{equation}
	with $\abs{\epsilon f(t)} \ll 1$ and $\epsilon = \hbar J / E_J$, where $E_J$ is the Josephson energy of each Josephson junction and $\phi_0 = \hbar/2e$ is the reduced flux quantum. It is also necessary to thread the inductive loop of the circuit with an external flux in the opposite direction and with half the amplitude as \cref{eqn:phix} to ensure the HO potential remains static~\cite{SM,You2019,Riwar2022,Bryon2023}. Here we have assumed the Josephson energies to be equal for simplicity, but we find the squeezing (logical fidelity) of the GKP Floquet states to decrease by less than 0.7~dB ($4\times10^{-3}$) for asymmetries of $\pm 0.05 E_J$ (see the SM~\cite{SM} for more details).

	As a concrete example, consider a HO frequency $\omega_0/2\pi = 1$~GHz and Josephson energy $E_J/h = 2$~GHz with $\epsilon = 1.25\times 10^{-3}$, which yields a ratio of $J/\omega_0 = 2.5 \times 10^{-3}$. For modest $N$, this is a very small modulation of the overall applied field, where $\phi_\text{e}(t)$ varies by no more than $0.01N\phi_0$ [\cref{eqn:phix}], and satisfies $\abs{\epsilon f(t)} \ll 1$. Since the external flux must be modulated according to \cref{eqn:f}, the maximum modulation frequency is $4N\omega_0$. Therefore, the limiting factor for generating highly squeezed GKP states is the maximum modulation frequency that can be achieved in an experiment. For an $\omega_0/2\pi=1$~GHz HO frequency, $N=4$ harmonics implies a maximum modulation frequency of $16$~GHz.
	
	\textit{State preparation.---}We now explain how the GKP Floquet states can be prepared. When the driving frequency is detuned from the HO frequency, the Floquet operator becomes approximately diagonal in the number basis, with nearly HO eigenstates as its Floquet states (see the SM~\cite{SM} for the derivation). By adiabatically tuning the driving frequency into resonance with the HO frequency, the low-energy HO eigenstates are adiabatically evolved into the GKP Floquet states. Therefore, we replace $\omega_0$ in \cref{eqn:f} with a tunable driving frequency $\omega(t)$. To maximize adiabaticity, the initial driving frequency is chosen to be incommensurate with the driving frequency $\omega(0) = \omega_0/(1 - \pi\times10^{-2})$, and is tuned into resonance using a sigmoid-function ramp with a numerically optimized slope [\cref{fig:AdiabaticRamps}]. In the SM we give the the precise equation of the ramp as well as a more detailed explanation for why an incommensurate detuning is chosen.
	
	Using this protocol, we numerically observe that the $\ket*{\psi^{(N)}_+}$ ($\ket*{\psi^{(N)}_-}$) Floquet state can be prepared from the $\ket{0}$ ($\ket{2}$) eigenstate of the HO. Note that the $\ket{0}$ and $\ket*{\psi^{(N)}_+}$ ($\ket{2}$ and $\ket*{\psi^{(N)}_-}$) states both have eigenvalue $+1$ ($-1$) for the Fourier transform operator since the adiabatic process preserves rotation symmetry. Furthermore, an arbitrary superposition $\alpha \ket{0} + \beta \ket{2}$ evolves into  $\alpha \ket*{\psi_+^{(N)}} + e^{i\phi}\beta \ket*{\psi_-^{(N)}}$, where $\phi$ is a phase factor that accounts for the dynamical phase difference acquired by the two Floquet states throughout the adiabatic evolution, thus allowing an arbitrary logical state to be prepared. In the following we focus on preparation of the $\ket*{\psi^{(4)}_+}$ Floquet state, and refer the reader to the SM~\cite{SM} for results on preparation of other states.
	
\begin{figure}[t]
		\subfloat{\label{fig:tfvsS}}
		\subfloat{\label{fig:tfvsFlog}}
		\centering
		\includegraphics[width=\linewidth]{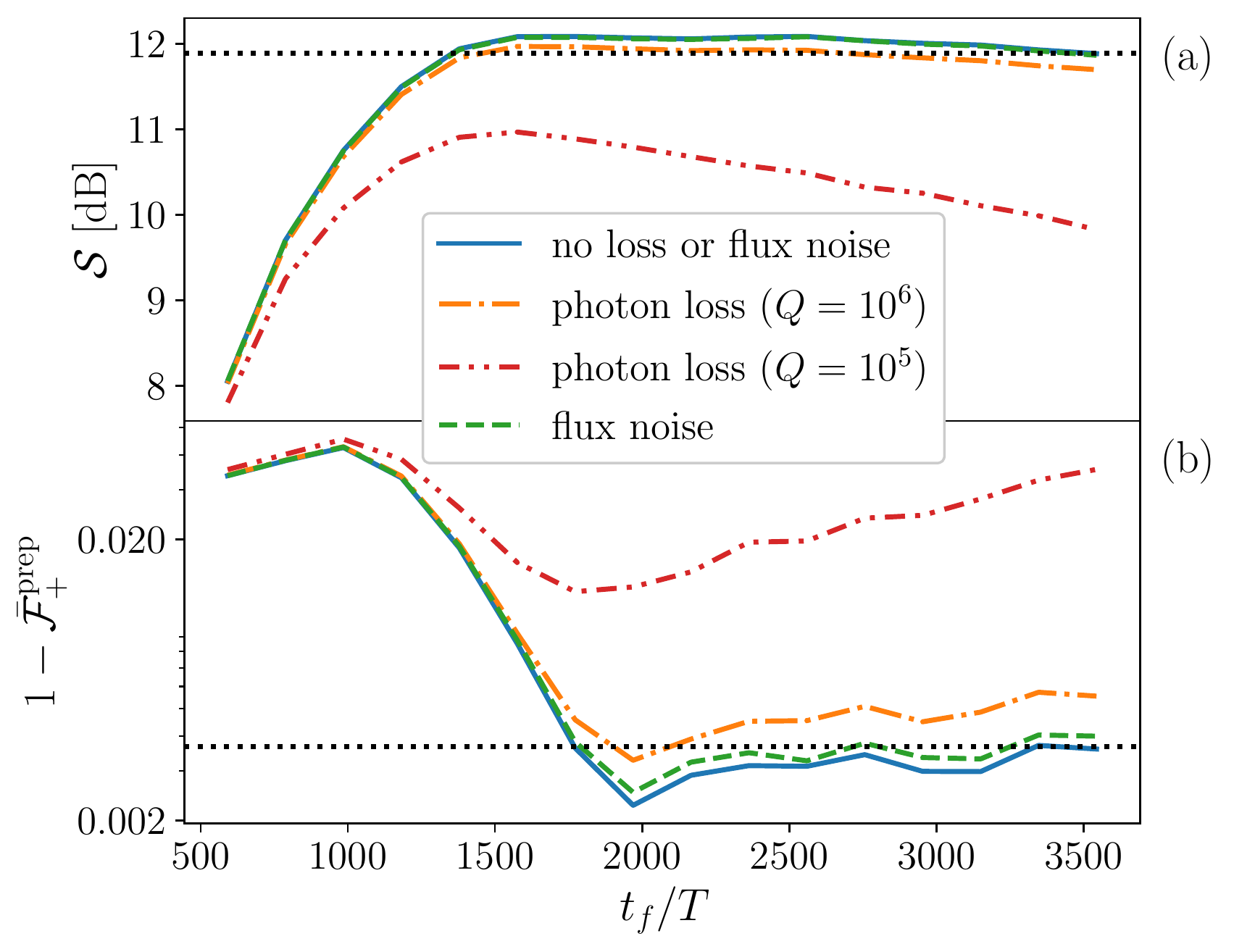}
		\caption{Quality of prepared GKP states. Squeezing (a) and logical infidelity (b) of the prepared GKP state as a function of preparation time. Blue solid lines denote preparation with no photon loss and no flux noise, orange (red) (dot-)dot-dashed lines represent the preparation with photon loss at a quality factor of $Q = \omega_0/\kappa = 10^6$ ($10^5$), and green dashed lines represent preparation with flux noise. The black dotted lines represent the values for the $\ket*{\psi_+^{(4)}}$ Floquet state [blue downward triangles in \cref{fig:NvsS,fig:NvsFlog}]. All results are for $J/\omega_0 = 2.5 \times 10^{-3}$ and $N=4$.}
		\label{fig:tfvsSFlog}
	\end{figure}

	For the remaining numerical results, we set $N=4$ and $J/\omega_0 = 2.5 \times 10^{-3}$. To assess the quality of the prepared state, $\rhoh(t_f)$, we compute its squeezing and logical fidelity \{$\bar{\mathcal{F}}_+^\text{prep} = \bra{H_+} \mathcal{D}\left[\rhoh(t_f)\right] \ket{H_+}$\} for different preparation times $t_f$ (\cref{fig:tfvsSFlog}). We observe both the squeezing and logical fidelity to increase over the range of preparation times of $1000 \lesssim t_f/T \lesssim 2000$ (blue solid lines in \cref{fig:tfvsSFlog}), after which they plateau and converge to the values for $\ket*{\psi^{(4)}_+}$ (dotted black lines in \cref{fig:tfvsSFlog}), consistent with the adiabatic theorem (in the SM~\cite{SM} we also plot the direct fidelity between the prepared state and the Floquet state). Interestingly, both the squeezing and logical fidelity of the prepared state reach higher values than for the true Floquet state $\ket*{\psi^{(4)}_+}$ before converging to their values, 
    and suggest an optimal preparation time of $t_f/T \approx 2000$.
	
	\textit{Noisy state preparation.---}In a closed environment, the adiabatic theorem dictates that the Floquet states may be prepared from the HO eigenstates with perfect fidelity by increasing the preparation time indefinitely. However, in the presence of decoherence processes, a longer preparation time may degrade the quality of the prepared state. Here we consider two decoherence processes: photon loss and dephasing due to flux noise. 
	
	To model photon loss, we consider the system Hamiltonian in \cref{eqn:H} coupled to a zero-temperature bath, described by the Lindblad master equation~\cite{Breuer2002}
	\begin{equation}
		\dot{\rhoh}(t) = -\frac{i}{\hbar}\bigl[\H(t),\, \rhoh(t)\bigr] + \kappa \Bigl( \a \rhoh(t) \a^\dag - \frac{1}{2}\bigl\{\a^\dag \a, \rhoh(t)\bigr\}\Bigr), \label{eqn:ME}
	\end{equation}
	where $\rhoh(t)$ is the reduced density matrix for the system at time $t$ and $\kappa$ is the photon-loss rate. We simulate photon loss using Monte Carlo sampling~\cite{Plenio1998} to generate the density matrix as an ensemble average of many quantum trajectories (see the SM~\cite{SM} for more details).
	
    In~\cref{fig:tfvsSFlog} we consider quality factors of $Q = \omega_0/\kappa = 10^6$ ($10^5$), which correspond to $T_1$ times of $1/\kappa \approx 160~\mu$s ($16~\mu$s) for an $\omega_0/2\pi=1$~GHz HO. At a quality factor of $Q = 10^5$, our numerical results (red dot-dot-dashed lines) reveal an optimal preparation time region at $1600 \lesssim t_f/T \lesssim 1800$, with a maximum squeezing (minimum logical infidelity) of $11.0$~dB ($1.3\times 10^{-2}$) achieved at the lower (upper) bound of this range. Increasing the quality factor to $Q=10^6$ improves these values (orange dot-dashed lines), with a maximum squeezing (minimum logical infidelity) of $12.0$~dB ($3.2 \times 10^{-3}$) attained at a preparation time of $t_f/T \approx 1600$ ($t_f/T \approx 2000$).
    
	In addition to photon loss, state preparation may be affected by noise from the external flux, which is used to modulate $f(t)$ in \cref{eqn:H}. However, when considering combined $1/f$ and white noise at experimentally measured noise amplitudes~\cite{Bylander2011,Stern2014,OMalley2015,Hutchings2017} (green dashed lines in \cref{fig:tfvsSFlog}), we find flux noise to have negligible impact on the quality of the prepared state, with a decrease of less than 0.02~dB ($6 \times 10^{-4}$) in the squeezing (logical fidelity) relative to the lossless case over all preparation times $t_f/T \lesssim 3500$. Thus, at quality factors of $Q \le 10^6$, photon loss will be the dominant source of decoherence for state preparation. In the SM~\cite{SM} we provide details on how this flux noise was modeled, as well as results at different flux noise strengths. 
 
    Overall, our results show that with flux noise and a quality factor of $Q=10^6$ ($Q=10^5$), a GKP state with squeezing $>11.9$~dB ($10.8$~dB) and logical infidelity $<4\times 10^{-3}$ ($2\times 10^{-2}$) may be prepared with a preparation time of $t_f/T \approx 1800$ (2000), which corresponds to $1.8~\mu$s ($2.0~\mu$s) for an $\omega_0/2\pi = 1$~GHz HO.
    
	\textit{Discussion and outlook.---}In this Letter we introduced a time-periodic Hamiltonian that has GKP states as its Floquet states, and showed how these states may be prepared by adiabatic evolution of low-energy HO eigenstates. We showed that this scheme may be realized in a superconducting circuit with a SQUID shunted by a linear superinductor and a capacitor. Numerical analysis of the impact of photon loss and flux noise on the squeezing and logical fidelity of the prepared state revealed that high-quality GKP states may be prepared on a microsecond timescale with experimentally achievable quality factors and typical flux noise amplitudes~\cite{Masluk2012,Kuzmin2019,Kuzmin2021,Peruzzo2020,Bylander2011,Stern2014,OMalley2015,Hutchings2017}. This represents an efficient way to prepare GKP states without the need for an ancilla qubit. As our protocol naturally gives rise to GKP magic states, universality can be achieved using only Clifford gates, which are realized with Gaussian bosonic operations on GKP states~\cite{Gottesman2001,Shaw2024a}.
	
    A natural question is whether the periodic driving may \textit{also} be used to stabilize the GKP state once it has been prepared, providing a form of autonomous error correction. Because of the Floquet heating phenomenon~\cite{Mori2023,Wang2023}, we find that the periodic drive decreases the lifetime of the GKP Floquet states (see the SM~\cite{SM} for further details). Thus, once the states are prepared, the drive should be turned off, and active error correction applied to stabilize the GKP states. This can be achieved via coupling to an ancilliary qubit, as has been done experimentally~\cite{CampagneIbarcq2020,Sivak2023}, or by interacting GKP modes, as has been studied in many theoretical works~\cite{Vuillot2019,Noh2020,Grimsmo2021,Noh2022}.
    
    Our work also opens opportunities for realizing more general Harper models beyond the GKP Hamiltonian, using simple superconducting circuits. Besides the rich and fascinating physics of the Harper model and its topological features, an intriguing avenue for future work is to investigate this broader class of models in the light of quantum error correction, and the potential to realize new quantum error correction schemes via periodic driving.

    \begin{acknowledgments}
    This work was supported by the Australian Research Council Centre of Excellence for Engineered Quantum Systems (EQUS, CE170100009). X.C.K. is also supported by an Australian Government Research Training Program (RTP) Scholarship. X.C.K. acknowledges access to the University of Sydney's high performance computing facility, Artemis, for obtaining numerical results. R.W.B. additionally acknowledges the support provided by the Deanship of Research Oversight and Coordination (DROC) at King Fahd University of Petroleum \& Minerals (KFUPM) through project No.~EC221010.
    \end{acknowledgments}
    
	\bibliography{refs}

\ifarXiv


\section*{Supplementary material}

In this Supplemental Material we provide supporting derivations and explanations for the letter "Gottesman-Kitaev-Preskill state preparation using periodic driving". In Sections S1 A, S4 C and S9 we use the convention that $\hbar = 1$ to increase the clarity of notation, and in the other sections we retain $\hbar$ in our notation. All equations, figures and sections in this Supplemental Material are labelled with an 'S', references to equations and figures without an 'S' refer to those in the main text.

## S1. FLOQUET ANALYSIS OF THE HARMONIC DRIVING SCHEME

In this section we provide a Floquet analysis of the harmonic driving scheme. In Section S1 A we utilize the Floquet-Magnus expansion to derive the first-order term for the effective Hamiltonian with and without a detuned driving frequency, and provide a bound on its second-order correction. In Section S1 B we numerically assess the impact of this bound on the GKP Floquet states of the harmonic driving scheme. In Section S1 A we set $\hbar = 1$.

### A. Floquet-Magnus expansion for the time-evolution operator

The Floquet theorem allows one to decompose the time-evolution operator $\hat{U}(t)$ for a periodic system into a time-independent part and a time-periodic part [S2]. The formulation we make use of is $\hat{U}(t) = e^{-i\hat{K}(t)}e^{-i\hat{H}_F t}$, where $\hat{H}_F$ is the effective Hamiltonian, which describes the long-term dynamics (with respect to a driving period), and $\hat{K}(t) = \hat{K}(t+T)$ is a periodic Hermitian operator called the micromotion operator, which describes the short-term dynamics, and satisfies the initial condition $\hat{K}(0) = 0$ [S3].

The Floquet-Magnus expansion is a series expansion for the effective Hamiltonian $\hat{H}_F = \hat{H}_F^{(1)} + \hat{H}_F^{(2)} + \ldots$ and the micromotion operator $\hat{K}(t) = \hat{K}^{(1)}(t) + \hat{K}^{(2)}(t) + \ldots$ in orders of the driving frequency $\omega$. The first-order terms are given by [S4]

$$\hat{H}_F^{(1)} = \frac{1}{T}\int_0^T dt'\, \hat{H}(t'), \tag{S1a}$$

$$\hat{K}^{(1)}(t) = \int_0^t dt'\, \hat{H}(t') - t\hat{H}_F^{(1)}. \tag{S1b}$$

Note that $\hat{H}_F^{(1)}$ corresponds to the time-average of the Hamiltonian over a single period.

*First-order term*

We now derive the effective Hamiltonian for the harmonic driving scheme. First we rewrite the GKP Hamiltonian and the time-dependent Hamiltonian in terms of displacement operators,

$$\hat{H}_{\text{GKP}} = -\frac{J}{2}\left(\hat{D}(\sqrt{2\pi}) + \hat{D}(-\sqrt{2\pi}) + \hat{D}(i\sqrt{2\pi}) + \hat{D}(-i\sqrt{2\pi})\right), \tag{S2a}$$

$$\hat{H}(t) = \omega_0 \hat{a}^\dagger \hat{a} - \frac{J}{2}f(t)\left(\hat{D}(i\sqrt{2\pi}) + \hat{D}(-i\sqrt{2\pi})\right). \tag{S2b}$$

Next, we move to a frame rotating at the HO frequency so that the time-evolution operator may be written as

$$\hat{U}(t) = e^{i\omega_0 \hat{a}^\dagger \hat{a} t}\hat{\mathcal{T}}e^{-i\int_0^t dt'\, \hat{\tilde{H}}(t')}, \tag{S3}$$

where $\hat{\mathcal{T}}$ denotes the time-ordering operator and $\hat{\tilde{H}}(t)$ is the time-dependent Hamiltonian in the rotating frame,

$$\hat{\tilde{H}}(t) = -\frac{J}{2}f(t)e^{i\omega_0 \hat{a}^\dagger \hat{a} t}\left(\hat{D}(i\sqrt{2\pi}) + \hat{D}(-i\sqrt{2\pi})\right)e^{-i\omega_0 \hat{a}^\dagger \hat{a} t} \tag{S4}$$

$$= -\frac{J}{2}f(t)\left(\hat{D}(i\sqrt{2\pi}e^{i\omega_0 t}) + \hat{D}(-i\sqrt{2\pi}e^{i\omega_0 t})\right). \tag{S5}$$

Here we used that $e^{i\theta \hat{a}^\dagger \hat{a}}\hat{D}(\alpha)e^{-i\theta \hat{a}^\dagger \hat{a}} = \hat{D}(e^{i\theta}\alpha)$.

To show that the time-averaged Hamiltonian is a truncation of the GKP Hamiltonian up to the $\pm 4N$ diagonals in the Fock basis, we write out the matrix elements in the Fock basis for both Hamiltonians,

$$\langle n| \hat{H}_{\text{GKP}} |n+k\rangle = -\frac{J}{2}\left(1 + e^{ik\pi/2} + e^{ik\pi} + e^{ik3\pi/2}\right) \langle n| \hat{D}(\sqrt{2\pi}) |n+k\rangle, \tag{S6a}$$

$$\langle n| \hat{\tilde{H}}(t) |n+k\rangle = -\frac{J}{2} f(t) e^{-ik\omega_0 t}\left(e^{ik\pi/2} + e^{ik3\pi/2}\right) \langle n| \hat{D}(\sqrt{2\pi}) |n+k\rangle, \tag{S6b}$$

where we used that $\langle m| \hat{D}(|\alpha|e^{i\arg\alpha}) |n\rangle = e^{i(m-n)\arg\alpha} \langle m| \hat{D}(|\alpha|) |n\rangle$ to extract the phases of the displacement operator arguments. Here, $k \in \mathbb{Z}$ labels the diagonals in the Fock basis as $n$ ranges over all integers for which $n+k \geq 0$. Note that $(1 + e^{ik\pi/2} + e^{ik\pi} + e^{ik3\pi/2})$ is non-zero only when $k = 0 \bmod 4$, whilst $(e^{ik\pi/2} + e^{ik3\pi/2})$ is non-zero only when $k = 0 \bmod 2$. This is a consequence of the fact that $\hat{H}_{\text{GKP}}$ and $\hat{\tilde{H}}(t)$ have $\mathbb{Z}_4$ and $\mathbb{Z}_2$ rotation symmetries, respectively. That is, they commute with the Fourier transform and parity operators $e^{i\pi\hat{a}^\dagger\hat{a}/2}$ and $e^{i\pi\hat{a}^\dagger\hat{a}}$, respectively.

From Eq. (S6b), we compute the matrix elements of the time-averaged Hamiltonian in the rotating frame [which is the first-order term for the effective Hamiltonian in the rotating frame via Eq. (S1a)] to be

$$\langle n| \hat{\tilde{H}}_F^{(1)} |n+k\rangle = -\frac{J}{2}\left(e^{ik\pi/2} + e^{ik3\pi/2}\right) \langle n| \hat{D}(\sqrt{2\pi}) |n+k\rangle \frac{1}{T}\int_0^T dt\, f(t) e^{-ik\omega_0 t}. \tag{S7}$$

We note that the integral in Eq. (S7) is the Fourier transform of the driving function when the HO frequency is equal to the driving function frequency ($\omega_0 = 2\pi/T$). If we set $f(t) = e^{ik'\omega_0 t}$, then we obtain the Kronecker delta function $\delta_{k,k'}$ from the integral. Thus, $\hat{\tilde{H}}_F^{(1)}$ will be non-zero, in general, along the $k'$-th diagonal in the Fock basis and zero elsewhere. Hence, by setting $f(t) = 2 + 4\sum_{k'=1}^N \cos(4k'\omega_0 t) = 2\sum_{k'=-N}^N e^{i4k'\omega_0 t}$ as in Eq. (4) in the main text, we obtain

$$\langle n| \hat{\tilde{H}}_F^{(1)} |n+k\rangle = -2J \langle n| \hat{D}(\sqrt{2\pi}) |n+k\rangle \delta_{k=0 \bmod 4}, \tag{S8}$$

for all $0 \leq k \leq 4N$.

Comparing with Eq. (S6a), we see that $\langle n| \hat{\tilde{H}}_F^{(1)} |n+k\rangle = \langle n| \hat{H}_{\text{GKP}} |n+k\rangle$ for all $0 \leq k \leq 4N$. Since $\hat{\tilde{H}}_F^{(1)}$ and $\hat{H}_{\text{GKP}}$ are Hermitian, we also have that $\langle n+k| \hat{\tilde{H}}_F^{(1)} |n\rangle = \langle n+k| \hat{H}_{\text{GKP}} |n\rangle$ for all $0 \leq k \leq 4N$. That is, the first-order term for the effective Hamiltonian in the rotating frame is equal to the GKP Hamiltonian up to the $\pm 4N$ diagonals in the Fock basis,

$$\hat{\tilde{H}}_F^{(1)} = \hat{H}_{\text{GKP}}^{(N)} := \sum_{n\in\mathbb{N}}\left[\langle n| \hat{H}_{\text{GKP}} |n\rangle |n\rangle\langle n| + \sum_{k=1}^{4N}\left(\langle n| \hat{H}_{\text{GKP}} |n+k\rangle |n\rangle\langle n+k| + \text{h.c.}\right)\right]. \tag{S9}$$

Moreover, when the HO frequency and driving frequency are equal, the effective Hamiltonian for $\hat{H}(t)$ is the same in the rotating frame ($\hat{H}_F = \hat{\tilde{H}}_F$). This is because we may absorb $e^{i\omega_0 \hat{a}^\dagger\hat{a}t}$ into $\hat{K}(t)$ by using the Floquet theorem in Eq. (S3) to write the time-evolution operator as

$$\hat{U}(t) = e^{i\omega_0 \hat{a}^\dagger \hat{a} t} e^{-i\hat{\tilde{K}}(t)} e^{-i\hat{\tilde{H}}_F t}, \tag{S10}$$

where $\hat{\tilde{K}}(t) = \hat{\tilde{K}}(t+T)$ is the micromotion operator in the rotating frame. Then, we have $e^{-i\hat{K}(t)} = e^{i\omega_0 \hat{a}^\dagger \hat{a} t} e^{-i\hat{\tilde{K}}(t)}$, which satisfies $e^{-i\hat{K}(t)} = e^{-i\hat{K}(t+T)}$ when $\omega_0 = 2\pi/T$.

*Higher-order corrections*

The second-order correction to the effective Hamiltonian is [S4]

$$\hat{H}_F^{(2)} = \frac{1}{2iT}\int_0^T dt \int_0^t dt'\, [\hat{H}(t), \hat{H}(t')], \tag{S11}$$

and the $n$-th order correction involves $n-1$ nested commutators of $\hat{H}(t)$. This second-order correction scales as $\|\hat{H}(t)\|^2 T$, where $\|\hat{H}(t)\|$ is some norm of the matrix over all times $t \in [0, T]$. From Eq. (S5), for fixed and finite $N$,

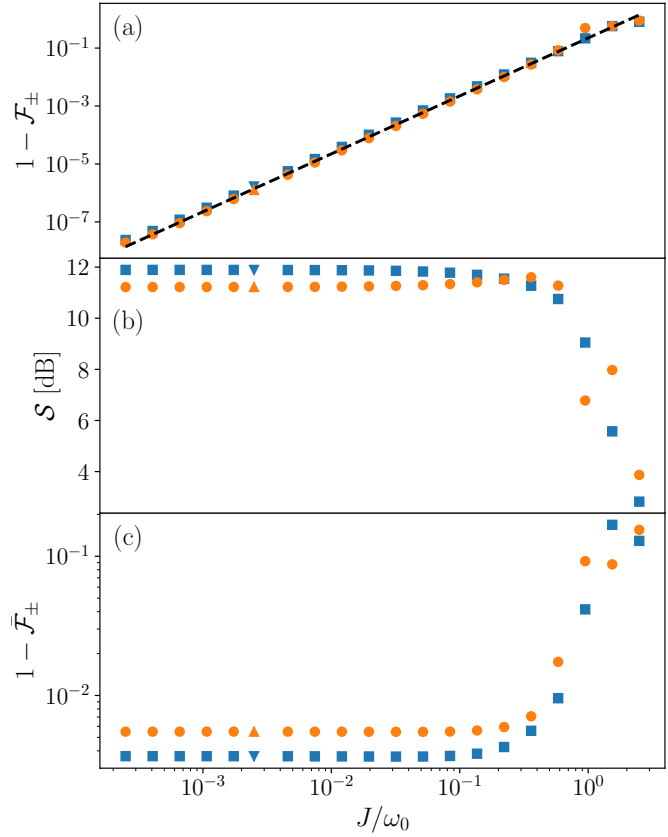

FIG. S1. Impact of $J/\omega_0$ on the quality of the GKP Floquet states. (a) Minimum infidelity of the eigenstates of $\hat{H}_{\text{GKP}}^{(4)}$ with the GKP Floquet states $|\psi_{\pm}^{(4)}\rangle$ for an $N=4$ harmonic driving scheme as a function of $J/\omega_0$. The black dashed line is a quadratic fit to the infidelity. (b) [(c)] Squeezing (logical infidelity) of the $|\psi_{\pm}^{(4)}\rangle$ states as a function of $J/\omega_0$. In all panels, blue squares (orange circles) denote values for the $|\psi_+^{(4)}\rangle$ ($|\psi_-^{(4)}\rangle$) state, and the blue downward (orange upward) triangles denote values for this state at $J/\omega_0 = 2.5 \times 10^{-3}$, as used in the main text.

we have that $\|\hat{\tilde{H}}(t)\| \sim J$ and thus when $T = 2\pi/\omega_0$, the second-order correction to $\hat{\tilde{H}}_F = \hat{H}_F$ scales as $J^2/\omega_0$. In general, the $n$-th order correction scales as $J^n/\omega_0^{n-1}$ for $n \geq 1$, hence the correction term in Eq. (5) in the main text.

In the case that $N \to \infty$, then $f(t)$ and hence $\|\hat{\tilde{H}}(t)\|$ becomes unbounded. However, in this case the Hamiltonian commutes with itself at all times. Therefore, the higher-order corrections to $\hat{\tilde{H}}_F = \hat{H}_F$ vanish in the case that $N \to \infty$, and we are left with $\hat{H}_F = \lim_{N \to \infty} \hat{H}_{\text{GKP}}^{(N)} = \hat{H}_{\text{GKP}}$, in agreement with the kicked HO model.

*Detuned driving frequency*

If we instead set $f(t) = e^{ik'\omega t}$ with $\omega/\omega_0 \in \mathbb{R} \setminus \mathbb{Q}$ in Eq. (S7) then the integral vanishes for all $k \neq k'$. Moreover, the effective Hamiltonian becomes $\hat{H}_F \approx \hat{\tilde{H}}_F + \delta\omega \hat{a}^\dagger \hat{a}$, where $\delta\omega = \omega - \omega_0$, because from Eq. (S10) the Floquet operator may be written as

$$\hat{U}_T = \hat{U}(T) = e^{i\omega \hat{a}^\dagger \hat{a} T} e^{-i\delta\omega \hat{a}^\dagger \hat{a} T} e^{-i\hat{\tilde{H}}_F T} \tag{S12}$$

$$\approx e^{-i(\hat{\tilde{H}}_F + \delta\omega \hat{a}^\dagger \hat{a})T}, \tag{S13}$$

where the second line follows from the first because $e^{i\omega \hat{a}^\dagger \hat{a} T} = \hat{I}$ and $\hat{\tilde{H}}_F$ is approximately diagonal in the Fock basis. Thus, a detuned, incommensurate driving frequency results in an effective Hamiltonian that is diagonal in the Fock basis to zeroth order in $J^2/\omega_0$. This explains why the HO eigenstates are approximate Floquet states of the harmonic driving function at the beginning of the adiabatic protocol when $\omega_0/\omega(0) = 1 - \pi \times 10^{-2} \in \mathbb{R} \setminus \mathbb{Q}$.

## B. Impact of $J/\omega_0$ on the GKP Floquet states

We now show how the ratio of $J/\omega_0$ impacts on the properties of the GKP Floquet states. Here we fix the number of harmonics to be $N = 4$, and vary the ratio $J/\omega_0$. Consistent with the correction term in Eq. (5) in the main text, we find that the maximum fidelity of the eigenstates of $\hat{H}_{\text{GKP}}^{(4)}$ with the $|\psi_\pm^{(4)}\rangle$ Floquet states scales quadratically with $J/\omega_0$ [see the agreement with the quadratic fit shown in Fig. S1(a)]. Specifically, our fidelity metric is $\mathcal{F}_\pm = \max_i |\langle \phi_i | \psi_\pm^{(4)} \rangle|^2$, where $|\phi_i\rangle$ are the eigenstates of $\hat{H}_{\text{GKP}}^{(4)}$. To assess the quality of the Floquet states when varying $J/\omega_0$, we compute their squeezing and logical fidelity and find that they retain squeezing (logical infidelity) of $> 11$ dB ($< 10^{-2}$) up to values of $J/\omega_0$ as large as 0.58 [Figs. S1(b) and S1(c)]. Beyond this point, their squeezing and logical fidelity decreases significantly, and their infidelity with the eigenstates of $\hat{H}_{\text{GKP}}^{(4)}$ becomes larger than 0.1. In all other numerical results, we focus on a value of $J/\omega_0$ that is well below 0.58 ($J/\omega_0 = 2.5 \times 10^{-3}$) since this results in a smaller amplitude of the non-linear modulation, which is easier to achieve in a circuit QED implementation (see the circuit QED implementation section in the main text as well as Section S3 here).

## S2. SUBSYSTEM DECOMPOSITION DECODER

In this section we explain how the subsystem decomposition decoder introduced in Ref. [S5] was implemented. For a complete description of this decoder we refer the reader to [S5], as well as [S6] for the original proposal of a bosonic subsystem decoder. The main idea is that the entire bosonic Hilbert space may be written as a tensor product of a logical subsystem $\mathcal{L}$ (containing the logical information) and a stabilizer subsystem $\mathcal{S}$ (containing the outcomes of stabilizer measurements). Performing a round of ideal error correction then corresponds to a completely-positive trace-preserving map $\mathcal{D} : \mathcal{L} \otimes \mathcal{S} \to \mathcal{L}$.

The entire Hilbert space may be written in the Zak basis [S7]

$$|k_1, k_2\rangle_\mathcal{Z} = \frac{1}{\sqrt[4]{\pi}} e^{ik_1 k_2/2} \sum_{s \in \mathbb{Z}} e^{2i\sqrt{\pi} s k_2} |k_1 + 2s\sqrt{\pi}\rangle_x, \tag{S14}$$

where $|z\rangle_x$ denotes a position eigenstate with eigenvalue $z$, and $k_1 \in (-\sqrt{\pi}/2, 3\sqrt{\pi}/2]$ and $k_2 \in (-\sqrt{\pi}/2, \sqrt{\pi}/2]$. This basis satisfies completeness

$$\int_{-\sqrt{\pi}/2}^{3\sqrt{\pi}/2} dk_1 \int_{-\sqrt{\pi}/2}^{\sqrt{\pi}/2} dk_2 \, |k_1, k_2\rangle_\mathcal{Z} \langle k_1, k_2| = 1, \tag{S15}$$

and orthogonality $\langle k_1, k_2 | k_1', k_2' \rangle = \delta(k_1 - k_1')\delta(k_2 - k_2')$. Associating the "left half" of this space with the logical zero state and the "right half" with the logical one state facilitates the following subsystem decomposition,

$$|0\rangle_\mathcal{L} \otimes |k_1, k_2\rangle_\mathcal{S} = |k_1, k_2\rangle_\mathcal{Z}, \tag{S16a}$$

$$|1\rangle_\mathcal{L} \otimes |k_1, k_2\rangle_\mathcal{S} = e^{i\sqrt{\pi} k_2/2} |k_1 + \sqrt{\pi}, k_2\rangle_\mathcal{Z}, \tag{S16b}$$

where now $k_1, k_2 \in (-\sqrt{\pi}/2, \sqrt{\pi}/2]$. The decoding map is then described by taking the partial trace over the stabilizer subsystem

$$\mathcal{D}[\hat{\rho}] = \text{Tr}_\mathcal{S}[\hat{\rho}] = \int_{-\sqrt{\pi}/2}^{\sqrt{\pi}/2} dk_1 \int_{-\sqrt{\pi}/2}^{\sqrt{\pi}/2} dk_2 \, {}_\mathcal{S}\langle k_1, k_2 | \hat{\rho} | k_1, k_2 \rangle_\mathcal{S}, \tag{S17}$$

so that $\mathcal{D}$ maps density matrices in $\mathcal{L} \otimes \mathcal{S}$ to $2 \times 2$ density matrices in $\mathcal{L}$.

By expanding an arbitrary state $\hat{\rho} \in \mathcal{L} \otimes \mathcal{S}$ in the subsystem decomposition basis [Eq. (S16)] and then substituting the expressions for the Zak states [Eq. (S14)] into Eq. (S17), the following expressions for the decoded state's matrix

elements may be obtained.

$$_{\mathcal{L}}\langle 0|\mathcal{D}[\hat{\rho}]|0\rangle_{\mathcal{L}} = \sum_{s\in\mathbb{Z}} \int_{-\sqrt{\pi}/2}^{\sqrt{\pi}/2} dk\ _x\langle k+2s\sqrt{\pi}|\hat{\rho}|k+2s\sqrt{\pi}\rangle_x, \tag{S18a}$$

$$_{\mathcal{L}}\langle 1|\mathcal{D}[\hat{\rho}]|1\rangle_{\mathcal{L}} = \sum_{s\in\mathbb{Z}} \int_{-\sqrt{\pi}/2}^{\sqrt{\pi}/2} dk\ _x\langle k+(2s+1)\sqrt{\pi}|\hat{\rho}|k+(2s+1)\sqrt{\pi}\rangle_x, \tag{S18b}$$

$$_{\mathcal{L}}\langle 0|\mathcal{D}[\hat{\rho}]|1\rangle_{\mathcal{L}} = {}_{\mathcal{L}}\langle 1|\mathcal{D}[\hat{\rho}]|0\rangle_{\mathcal{L}}^* = \frac{1}{\pi}\sum_{s,s'\in\mathbb{Z}} \frac{(-1)^{s'-s}}{s'-s+1/2} \int_{-\sqrt{\pi}/2}^{\sqrt{\pi}/2} dk\ _x\langle k+2s\sqrt{\pi}|\hat{\rho}|k+(2s'+1)\sqrt{\pi}\rangle_x. \tag{S18c}$$

Since we work in the Fock basis for our numerical analysis, we pre-compute a tensor of integrals $M_{m,n}^{(\mu,\nu)}$ such that $\mathcal{D}[\hat{\rho}]$ may be computed in the Fock basis by

$$_{\mathcal{L}}\langle \mu|\mathcal{D}[\hat{\rho}]|\nu\rangle_{\mathcal{L}} = \sum_{m,n=0}^{\infty} M_{m,n}^{(\mu,\nu)} \langle m|\hat{\rho}|n\rangle. \tag{S19}$$

The elements of the tensor are given by

$$M_{m,n}^{(0,0)} = \frac{1}{\sqrt{\pi 2^{m+n}m!n!}} \sum_{s\in\mathbb{Z}} \int_{-\sqrt{\pi}/2}^{\sqrt{\pi}/2} dk\, e^{-(k+2s\sqrt{\pi})^2} H_m(k+2s\sqrt{\pi})H_n(k+2s\sqrt{\pi}), \tag{S20a}$$

$$M_{m,n}^{(1,1)} = \frac{1}{\sqrt{\pi 2^{m+n}m!n!}} \sum_{s\in\mathbb{Z}} \int_{-\sqrt{\pi}/2}^{\sqrt{\pi}/2} dk\, e^{-(k+(2s+1)\sqrt{\pi})^2} H_m(k+(2s+1)\sqrt{\pi})H_n(k+(2s+1)\sqrt{\pi}), \tag{S20b}$$

$$M_{m,n}^{(0,1)} = M_{m,n}^{(1,0)*} = \frac{1}{\pi\sqrt{\pi 2^{m+n}m!n!}} \sum_{s,s'\in\mathbb{Z}} \frac{(-1)^{s'-s}}{s'-s+1/2}$$
$$\times \int_{-\sqrt{\pi}/2}^{\sqrt{\pi}/2} dk\, e^{-\frac{(k+2s\sqrt{\pi})^2+(k+(2s'+1)\sqrt{\pi})^2}{2}} H_m(k+2s\sqrt{\pi})H_n(k+(2s'+1)\sqrt{\pi}). \tag{S20c}$$

Here, $H_n(x)$ denotes the physicist's Hermite polynomial, which arises from using the following formula for the position-Fock basis overlap [S8],

$$_x\langle x|n\rangle = \frac{e^{-x^2/2}H_n(x)}{\sqrt[4]{\pi}\sqrt{2^n n!}}. \tag{S21}$$

For numerical purposes, the infinite sums in Eq. (S20) must be truncated at a finite range $\pm S$. For the states considered in this letter, we find a truncation of $S = 7$ to be sufficient to achieve convergence in the logical fidelity.

An alternative implementation of the subsystem decomposition decoder may be constructed using logical Pauli measurement operators [S9]. This method still requires the truncation of an infinite sum. However, no integrals need to be computed numerically (for the square GKP code), which allows for a larger sum truncation to be computed accurately. We have verified that this alternative approach gives identical results, but reserve an explanation of the implementation for a future publication due to its use of analytical results that are beyond the scope of this letter.

## S3. CIRCUIT QED IMPLEMENTATION

In this section we provide additional details on the circuit QED implementation of the harmonic driving scheme. In Section S3 A we show how the Hamiltonian in Eq. (3) of the main text may be obtained from the circuit diagram in Fig. 2(a) of the main text, and in Sections S3 B and S3 C we analyze the impact of relaxing some of the assumptions in the derivation.

### A. Derivation of circuit Hamiltonian

We start by considering the most general form of the circuit shown in Fig. S2, where we allow for a potentially asymmetric SQUID with Josephson energies $E_{J1}$, $E_{J2}$ and capacitances $C_{J1}$, $C_{J2}$. Methods of circuit quantization

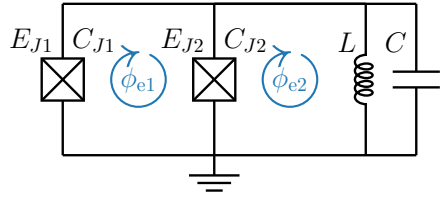

FIG. S2. General circuit diagram for the harmonic driving scheme. The circuit comprises a possibly asymmetric SQUID loop with Josephson energies $E_{J1}, E_{J2}$ and capacitances $C_{J1}, C_{J2}$ shunted by a linear inductor with inductance $L$ and capacitor with capacitance $C$. The two loops in the circuit are threaded by external fluxes $\phi_{e1}$ and $\phi_{e2}$.

in the presence of time-dependent external fluxes have recently been studied in Ref. [S10–S12], with experimental results in Ref. [S13] corroborating their theoretical predictions. These studies highlight the fact that the form of the Hamiltonian is gauge-dependent, and that a particular gauge choice, referred to as the *irrotational gauge*, must be made in order for the Hamiltonian to resemble the time-independent form. In this gauge, the Hamiltonian for the circuit in Fig. S2 takes the form

$$\hat{H} = 4E_C\hat{n}^2 + \frac{E_L}{2}\left(\hat{\varphi} - \frac{C_{J1}}{C_\Sigma}\varphi_{e1} - \frac{C_{J1}+C_{J2}}{C_\Sigma}\varphi_{e2}\right)^2$$
$$- E_{J1}\cos\left(\hat{\varphi} + \frac{C_{J2}+C}{C_\Sigma}\varphi_{e1} + \frac{C}{C_\Sigma}\varphi_{e2}\right) - E_{J2}\cos\left(\hat{\varphi} - \frac{C_{J1}}{C_\Sigma}\varphi_{e1} + \frac{C}{C_\Sigma}\varphi_{e2}\right). \quad \text{(S22)}$$

Here, $\hat{n}$ and $\hat{\varphi}$ are the conjugate variables for the single degree of freedom in the circuit satisfying $[\hat{\varphi}, \hat{n}] = i$; $\varphi_{ej} = \phi_{ej}/\phi_0$ is the phase associated with each of the external fluxes, with $\phi_0 = \hbar/2e$ the reduced flux quantum; $E_C = e^2/2C_\Sigma$ and $E_L = \phi_0^2/L$ are the capacitive and inductive energies, respectively; and $C_\Sigma = C_{J1}+C_{J2}+C$ is the total capacitance in the circuit. This Hamiltonian may be obtained by following the procedure outlined in Ref. [S10], which assumes the irrotational gauge choice is made, or by following the method for quantizing more general circuits using symplectic geometry introduced in Ref. [S12], and then making the irrotational gauge choice.

Setting $\varphi_{e2} = -\varphi_{e1}C_{J1}/(C_{J1}+C_{J2})$ allows us to eliminate all terms coupling the inductor to the external flux:

$$\hat{H} = 4E_C\hat{n}^2 + \frac{E_L}{2}\hat{\varphi}^2 - E_{J1}\cos\left(\hat{\varphi} + \frac{C_{J2}}{C_{J1}+C_{J2}}\varphi_e\right) - E_{J2}\cos\left(\hat{\varphi} - \frac{C_{J1}}{C_{J1}+C_{J2}}\varphi_e\right), \quad \text{(S23)}$$

where we have relabelled $\varphi_{e1} = \varphi_e$ as per our notation in Fig. 2(a) of the main text. For a symmetric SQUID with $E_{J1} = E_{J2} = E_J$ and $C_{J1} = C_{J2}$ the Hamiltonian reduces to

$$\hat{H} = 4E_C\hat{n}^2 + \frac{E_L}{2}\hat{\varphi}^2 - 2E_J\cos(\varphi_e/2)\cos(\hat{\varphi}), \quad \text{(S24)}$$

where we have utilized the trig identity $\cos(\alpha+\beta) = \cos\alpha\cos\beta - \sin\alpha\sin\beta$. By setting $E_C/\pi = 2\pi E_L = \hbar\omega_0/2$, this becomes

$$\hat{H} = \frac{\hbar\omega_0}{2}(\hat{x}^2 + \hat{p}^2) - 2E_J\cos(\varphi_e/2)\cos(2\sqrt{\pi}\hat{x}), \quad \text{(S25)}$$

where the position and momentum are defined by $\hat{x} = \hat{\varphi}/2\sqrt{\pi}$ and $\hat{p} = 2\sqrt{\pi}\hat{n}$, so as to satisfy $[\hat{x},\hat{p}] = i$. Note that fixing $E_C/\pi = 2\pi E_L$ is equivalent to the impedance constraint $Z = 2R_Q$, where $Z = \sqrt{L/C_\Sigma}$ is the impedance of the circuit and $R_Q = h/(2e)^2$ is the resistance quantum.

By modulating the external flux as $\varphi_e(t) = \pi - \epsilon f(t)$ with $|\epsilon f(t)| \ll 1$, we may invoke the small-angle approximation to write $\cos(\varphi_e/2) = \sin(\epsilon f(t)/2) \approx \epsilon f(t)/2$. The resulting Hamiltonian is then

$$\hat{H} = \hbar\omega_0 \hat{a}^\dagger \hat{a} - E_J \epsilon f(t)\cos(2\sqrt{\pi}\hat{x}), \quad \text{(S26)}$$

where we have ignored the constant energy shift $\hbar\omega_0/2$. Therefore, Eq. (3) in the main text is recovered with $\hbar J = E_J \epsilon$.

### B. Violation of the impedance condition

Violating the impedance constraint $Z = 2R_Q$ changes the coefficient of $\hat{x}$ in the cosine potential of the Hamiltonian in Eq. (3) of the main text [specifically, we have that $\hat{H}/\hbar = \omega_0 \hat{a}^\dagger \hat{a} - Jf(t)\cos(\sqrt{2\pi Z/R_Q}\hat{x})$]. This results in an

effective Hamiltonian that does not necessarily include the stabilizers of the GKP code, and thus GKP Floquet states are not guaranteed for arbitrary values of impedance. However, we find that for variations in the impedance within $2R_Q \pm 0.1 R_Q$, the squeezing (logical fidelity) of the Floquet states decreases by less than 0.7 dB ($4 \times 10^{-3}$) [Figs. S3(a) and S3(b)]. Our results also reveal that the $|\psi_-^{(4)}\rangle$ state is more robust to variations in the impedance, with a maximum decrease in the squeezing (logical fidelity) of less than 0.08 dB ($3.3 \times 10^{-4}$) for variations in impedance within $2R_Q \pm 0.1 R_Q$. Interestingly, the squeezing and logical fidelity for the $|\psi_+^{(4)}\rangle$ state is higher in the region $1.93 R_Q < Z < 2R_Q$. Therefore, the $|\psi_+^{(4)}\rangle$ state actually benefits from slightly lower values of impedance than $2R_Q$.

### C. Unequal Josephson energies

In our circuit derivation (see Section S3 A) we have assumed the two Josephson energies in the SQUID loop to be equal for simplicity. If instead we take the two Josephson energies to be unequal by an amount $\Delta E_J$, then the Hamiltonian for the circuit becomes [c.f. Eq. (S25)]

$$\hat{H} = \hbar\omega_0 \hat{a}^\dagger \hat{a} - 2E_J \cos(\varphi_\mathrm{e}/2)\cos(2\sqrt{\pi}\hat{x}) - \Delta E_J \sin(\varphi_\mathrm{e}/2)\sin(2\sqrt{\pi}\hat{x}), \tag{S27}$$

where $2E_J$ is the sum of the two Josephson energies. After fixing the external flux to be $\varphi_\mathrm{e} = \pi - \epsilon f(t)$, with $\hbar J = E_J \epsilon$ as before, to first order in $|\epsilon f(t)|$ the Hamiltonian becomes

$$\hat{H} = \hbar\omega_0 \hat{a}^\dagger \hat{a} - \hbar J f(t) \cos(2\sqrt{\pi}\hat{x}) - \Delta E_J \sin(2\sqrt{\pi}\hat{x}). \tag{S28}$$

Thus, the result of unequal Josephson energies is a spurious $\sin(2\sqrt{\pi}\hat{x})$ potential [c.f. Eq. (3) in the main text]. However, we find that for asymmetries of $|\Delta E_J| \leq 0.05 E_J$, the squeezing (logical fidelity) of the GKP Floquet states is reduced by less than 0.7 dB ($4 \times 10^{-3}$) by the $\sin(2\sqrt{\pi}\hat{x})$ potential [Figs. S3(c) and S3(d)]. Here we have fixed $\epsilon = 1.25 \times 10^{-3}$ as in the main text, which implies that $E_J = 2\hbar\omega_0$ when $J/\omega_0 = 2.5 \times 10^{-3}$. Similarly to perturbations in the impedance, we find the $|\psi_-^{(4)}\rangle$ state to be more robust to asymmetries in the SQUID loop than the $|\psi_+^{(4)}\rangle$ state. In contrast to perturbations in the impedance, neither Floquet state benefits from unequal Josephson energies.

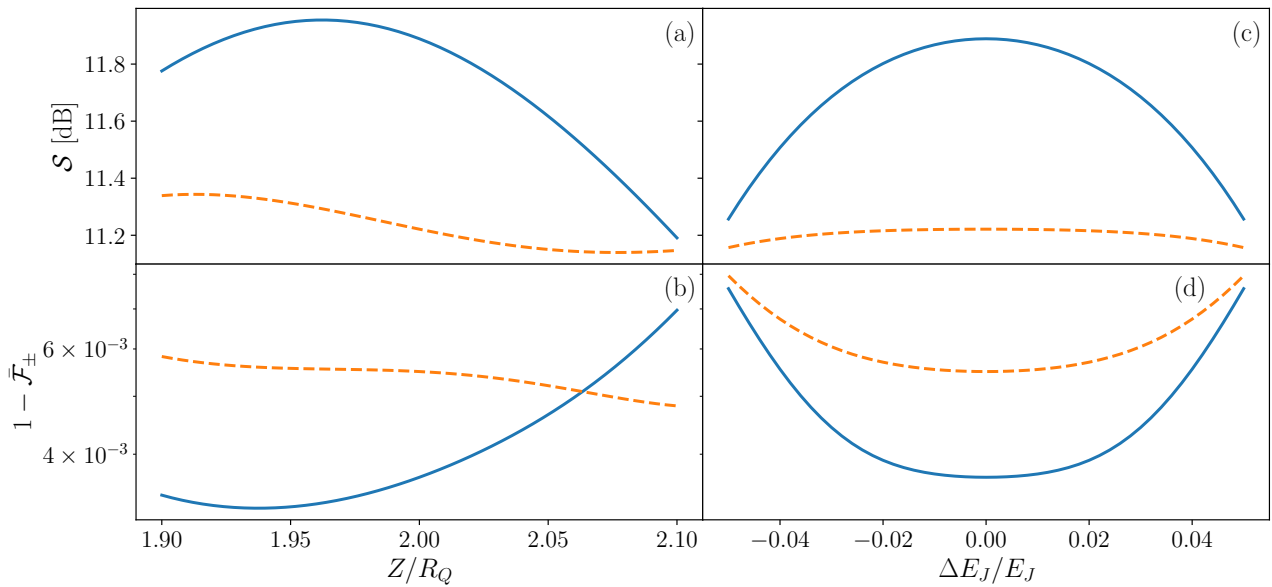

FIG. S3. Impact of variations in circuit parameters on the GKP Floquet states of the harmonic driving scheme. The first (second) row shows the squeezing (logical infidelity) and the first (second) column shows these as a function of impedance (Josephson energy asymmetry). Blue solid (orange dashed) lines denote values for the $|\psi_+^{(4)}\rangle$ ($|\psi_-^{(4)}\rangle$) state. All results are for an $N = 4$ harmonic driving scheme at $J/\omega_0 = 2.5 \times 10^{-3}$, with the HO frequency fixed at $\omega_0 = 1/\sqrt{LC_\Sigma}$. In (c) and (d) $\epsilon = 1.25 \times 10^{-3}$ so that $E_J = 2\hbar\omega_0$

## S4. ADIABATIC STATE PREPARATION

In this section we provide details on the adiabatic state preparation protocol. In Section S4 A we give the exact equations for the adiabatic ramps shown in Fig. 2(b) of the main text, in Section S4 B we give an alternative preparation scheme, and in Section S4 C we derive an adiabatic condition that applies to our protocol.

### A. Adiabatic frequency ramp

To tune the driving frequency we choose $K$ driving frequencies from the following function

$$\omega[k] = \frac{\omega_0 - \omega_1}{2}\,\text{erf}\bigl(\alpha(k - K/2)\bigr) + \frac{\omega_0 + \omega_1}{2}, \quad k \in \mathbb{N}, \tag{S29}$$

which represents an error-function ramp ($\text{erf}(x) = \frac{2}{\sqrt{\pi}} \int_0^x dt\, e^{-t^2}$) with slope $\alpha$, centered at $K/2$ and horizontal asymptotes at $\omega_0$ and $\omega_1 := \omega_0/(1 - \pi \times 10^{-2})$. The initial state is then evolved under each "Floquet operator" [S14] $\hat{U}_{T[k]}$ for a time $T[k] = 2\pi/\omega[k]$ so that the prepared state is given by

$$|\psi(t_f)\rangle = \prod_{k=1}^{K} \hat{U}_{T[k]} |\psi(0)\rangle, \tag{S30}$$

with a preparation time $t_f = \sum_{k=1}^{K} T[k]$. Equivalently, this may be defined as continuous evolution under the time-dependent Hamiltonian in Eq. (3) of the main text with $\omega(t) = \omega[k(t)]$, where $k(t) = \max\{l \in \mathbb{N} : \sum_{j \leq l} T[j] < t\}$, for a total time $t_f$. By numerically optimizing the squeezing and logical fidelity of the prepared state for a specific preparation time we find a slope of $\alpha = 5.75/K$ to be approximately optimal.

### B. Alternative adiabatic protocol

In Section S1 A we noted that the HO eigenstates are only *approximate* Floquet states of the detuned Floquet operator. It is also possible to ramp the amplitude of the cosine potential in Eq. (3) of the main text from 0 to $J$ by modifying the driving function to be

$$f(t) = A(t)\Bigl(2 + 4\sum_{n=1}^{N} \cos(4n\omega(t)t)\Bigr), \tag{S31}$$

so that initially the Hamiltonian is $\hat{H} = \hbar\omega_0 \hat{a}^\dagger \hat{a}$ whose eigenstates are precisely the HO eigenstates. Due to the fact that quasienergies are only defined modulo $\omega(t)$, all eigenstates of the HO are degenerate Floquet states if $\omega(t)$ and

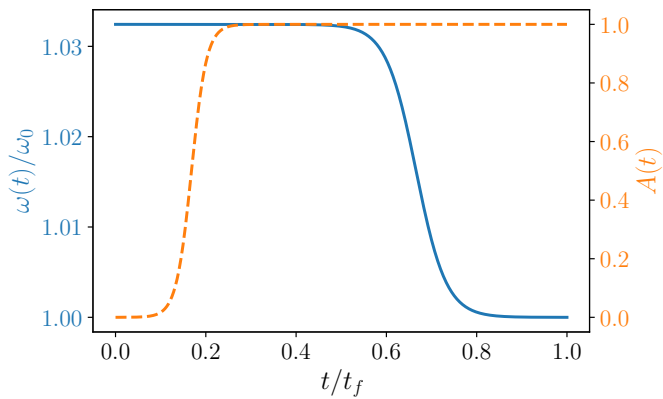

FIG. S4. Alternative adiabatic protocol. The driving frequency (amplitude) is shown by the solid blue (dashed orange) line. Both ramps are error function ramps. The initial driving frequency is $\omega(0) = \omega_0/(1 - \pi \times 10^{-2})$.

$\omega_0$ are commensurate. In the next section we show that adiabaticity cannot be maintained in this case. Therefore, the driving frequency must still be tuned from an initially incommensurate value, and then once the cosine potential has been turned on, the driving frequency can be tuned into resonance with the HO frequency.

In Fig. S4 we show a possible set of ramps for the driving amplitude $A(t)$ and driving frequency $\omega(t)$ in Eq. (S31) that also prepares GKP states from the HO eigenstates. However, we find that for equivalent preparation times, the diabatic errors incurred by the larger slopes necessary to accommodate both ramps outweighs the error incurred by beginning in an approximate eigenstate, the latter of which decreases with $J/\omega_0$ (see Sections S1 A and S1 B).

### C. Adiabatic condition

We are not aware of any literature that addresses the adiabatic theorem for an approximately periodic system with a discretely-modulated period. A proof of the adiabatic theorem in a Floquet setting (i.e., a truly periodic system) is presented in Ref. [S15]. Here we adapt the proof given in Ref. [S15] to our setting, allowing us to derive an adiabatic condition that motivates the protocols described here. For this section we set $\hbar = 1$.

We start by dividing the time evolution into discrete segments, defined by the driving periods, which are themselves functions of time due to the modulation of the driving frequency. To this end, let $\hat{U}_{T[k]} = \hat{\mathcal{T}} e^{-i \int_{T[k-1]}^{T[k]} dt' \hat{H}(t')}$ be the $k$-th Floquet operator during the adiabatic process, where $k = k(t) = \max\{l \in \mathbb{N} : \sum_{j \le l} T[j] < t\}$ is implicitly a function of time, and $T[k] = 2\pi/\omega[k]$ is the $k$-th driving period (e.g., with $\omega[k]$ defined by the adiabatic protocol given in Section S4 A). Let $s = t/t_f$ parameterize the time and $\Delta s[k] = T[k]/t_f$ be the $k$-th dimensionless driving period. Let $|\psi_n[k]\rangle$ be the Floquet states for the $k$-th Floquet operator with quasienergies $\varepsilon_n[k]$, and $|\psi(s)\rangle$ be the state that is evolving under $\hat{H}(s)$ [S16].

Since each of the Floquet operators is a unitary operator, by the spectral theorem, their eigenstates form a basis for the total Hilbert space. Therefore, we can write the time-evolved state at time $s$ as a superposition of the $k$-th Floquet states,

$$|\psi(s)\rangle = \sum_n c_n(s) e^{-i\theta_n[k]} |\psi_n[k]\rangle, \tag{S32}$$

where $c_n(s)$ are some time-dependent coefficients and $\theta_n[k] = \sum_{l=0}^{k-1} \varepsilon_n[l] T[l]$ is the dynamical phase accumulated by the $n$-th Floquet state at time $s$.

By the definition of $\hat{U}_{T[k]}$, we have that

$$|\psi(s + \Delta s[k])\rangle = \hat{U}_{T[k]} |\psi(s)\rangle. \tag{S33}$$

Substituting Eq. (S32) into Eq. (S33) and left multiplying by $\langle \psi_m[k]|$, a particular dual Floquet state of $\hat{U}_{T[k]}$, we have

$$\sum_n c_n(s + \Delta s[k]) e^{-i\theta_n[k+1]} \langle \psi_m[k]|\psi_n[k+1]\rangle = e^{-i\varepsilon_m[k]T[k]} e^{-i\theta_m[k]} c_m(s), \tag{S34}$$

where we used the orthogonality of the Floquet states together with the fact that $\langle \psi_m[k]| \hat{U}_{T[k]} = e^{-i\varepsilon_m[k]T[k]} \langle \psi_m[k]|$.

Now, by Taylor expansion, we can write the Floquet states of $\hat{U}_{T[k+1]}$ as $|\psi_n[k+1]\rangle = |\psi_n[k]\rangle + \Delta s[k] \frac{d}{ds} |\psi_n[k]\rangle$ to first-order in $\Delta s[k]$. So long as we do not follow a closed path in the space of time-dependent Hamiltonians we can choose the phase on the Floquet states at each time such that they satisfy the parallel transport condition [S17], which requires that the change in a Floquet state is orthogonal to the state itself ($\langle \psi_m[k]| \frac{d}{ds} |\psi_m[k]\rangle = 0$) [S18]. In this case Eq. (S34) becomes

$$c_m(s + \Delta s[k]) e^{-i\theta_m[k+1]} + \sum_{n \ne m} c_n(s + \Delta s[k]) e^{-i\theta_n[k+1]} \Delta s[k] \langle \psi_m[k]| \frac{d}{ds} |\psi_n[k]\rangle = e^{-i\theta_m[k+1]} c_m(s), \tag{S35}$$

where we have also combined the dynamical phases on the right-hand side of Eq. (S34).

By dividing both sides by $\Delta s[k] e^{-i\theta_m[k+1]}$ and rearranging, Eq. (S35) can be rewritten as

$$\frac{c_m(s + \Delta s[k]) - c_m(s)}{\Delta s[k]} = -\sum_{n \ne m} e^{-i(\theta_n[k+1] - \theta_m[k+1])} \langle \psi_m[k]| \frac{d}{ds} |\psi_n[k]\rangle c_n(s + \Delta s[k]). \tag{S36}$$

Taking the limit that $\Delta s[k] \to 0$ in Eq. (S36) (so that the total evolution time is much greater than any driving period), we arrive at the following differential equation for the time-dependent coefficients

$$\frac{d}{ds}c_m(s) = -\sum_{n\neq m} e^{-i(\theta_n[k]-\theta_m[k])} \langle\psi_m[k]|\frac{d}{ds}|\psi_n[k]\rangle c_n(s),$$

whose solution is given by

$$c_m(s) = c_m(0) - \sum_{n\neq m} \int_0^s ds' \, e^{-i(\theta_n[k']-\theta_m[k'])} \langle\psi_m[k']|\frac{d}{ds'}|\psi_n[k']\rangle c_n(s'), \tag{S37}$$

where $k' = k(s't_f)$. When the integral in Eq. (S37) vanishes, transitions between Floquet eigenstates do not occur so that adiabaticity is achieved.

To show that adiabaticity may be achieved by increasing the preparation time $t_f$, we proceed by expanding the integral in Eq. (S37) in orders of $1/t_f$. Since the dynamical phase accumulated by the Floquet states only changes at the beginning of each period, the minimal change in dynamical phase is $de^{i\theta_n[k]} = e^{i\theta_n[k+1]} - e^{i\theta_n[k]}$. In the limit that $\Delta s[k] \to 0$, we may therefore compute the derivative of the first term in the integrand of Eq. (S37) as

$$\frac{d}{ds}e^{-i(\theta_n[k]-\theta_m[k])} = \frac{e^{-i(\theta_n[k+1]-\theta_m[k+1])} - e^{-i(\theta_n[k]-\theta_m[k])}}{\Delta s[k]} \tag{S38}$$

$$= \frac{e^{-i(\theta_n[k]-\theta_m[k])}(e^{-i(\varepsilon_n[k]-\varepsilon_m[k])T[k]} - 1)}{T[k]/t_f}, \tag{S39}$$

where we used the definition of the dynamical phase and the dimensionless driving period in the second line. We may then use Eq. (S39) to evaluate the integral in Eq. (S37) by parts, yielding

$$\int_0^s ds' \, e^{-i(\theta_n[k']-\theta_m[k'])} \langle\psi_m[k']|\frac{d}{ds'}|\psi_n[k']\rangle c_n(s') = \frac{1}{t_f}\left[T[k']\frac{e^{-i(\theta_n[k']-\theta_m[k'])}}{e^{-i(\varepsilon_n[k']-\varepsilon_m[k'])T[k']} - 1} \langle\psi_m[k']|\frac{d}{ds'}|\psi_n[k']\rangle c_n(s')\right]_{s'=0}^{s'=s}$$

$$- \frac{1}{t_f}\int_0^s ds' \, e^{-i(\theta_n[k']-\theta_m[k'])} \frac{d}{ds'}\left[\frac{T[k']}{e^{-i(\varepsilon_n[k']-\varepsilon_m[k'])T[k']} - 1} \langle\psi_m[k']|\frac{d}{ds'}|\psi_n[k']\rangle c_n(s')\right]. \tag{S40}$$

Using Eq. (S39) again to integrate the second term in Eq. (S40) by parts, we find that it is suppressed by an additional factor of $1/t_f$ relative to the first term. Therefore, to first order in $1/t_f$, the time-evolution of the coefficients is given by

$$c_m(s) = c_m(0) - \frac{1}{t_f}\sum_{n\neq m}\left[T[k']\frac{e^{-i(\theta_n[k']-\theta_m[k'])}}{e^{-i(\varepsilon_n[k']-\varepsilon_m[k'])T[k']} - 1} \langle\psi_m[k']|\frac{d}{ds'}|\psi_n[k']\rangle c_n(s')\right]_{s'=0}^{s'=s}. \tag{S41}$$

When the terms inside the sum of Eq. (S41) are small compared with the preparation time, the coefficients do not evolve with time. Thus, when $c_n(0) = 0$ for all $n \neq m$ in Eq. (S32), the time-evolved state remains in an instantaneous Floquet state of the Floquet operator throughout the evolution. That is, the adiabatic condition may be written as

$$\frac{\left|\langle\psi_m|\frac{d}{ds}|\psi_n\rangle\right|}{|e^{-i\varepsilon_n T} - e^{-i\varepsilon_m T}|} \ll \frac{t_f}{T} \quad \forall n \neq m, \tag{S42}$$

where we have omitted the arguments for the Floquet states, quasienergies and the driving period for clarity, with the understanding that they are all defined at time $s$.

Eq. (S42) shows that adiabaticity is increased for longer total preparation time with respect to the driving period. Crucially, it also shows that adiabaticity may be lost when two Floquet states have equal quasienergies modulo $2\pi/T$ since the denominator on the left-hand side vanishes. This means that, if initially $\hat{H}(0) = \omega_0 \hat{a}^\dagger \hat{a}$ (as in the protocol described in Section S4 B) and thus $\varepsilon_n = n\omega_0$, the denominator vanishes for all $m$ and $n$ if the driving frequency is on resonance with the HO frequency, $\omega = \omega_0$. Similarly, if the driving frequency and HO frequency are commensurate, $\omega/\omega_0 = q \in \mathbb{Q}$, then the denominator vanishes for all $m = n$ mod $q$. This explains why the driving frequency must be made incommensurate with the HO frequency in the alternative adiabatic protocol, ensuring that the denominator in Eq. (S42) does not vanish in the early part of the protocol. Once $A(t)$ is turned on, the quasienergy spectrum has become sufficiently different from that of the HO, and at that point it is possible to adjust the driving frequency until it is equal to $\omega_0$ while remaining adiabatic.

## S5. ARBITRARY LOGICAL STATE PREPARATION

Here we give additional details on preparation of an arbitrary logical GKP state. To prepare an arbitrary state the same state preparation protocol discussed in the main text may be used by starting with a different initial state. We remind the reader that initializing in the $|0\rangle$ ($|2\rangle$) HO eigenstate prepares the $|\psi_+^{(N)}\rangle$ ($|\psi_-^{(N)}\rangle$) Floquet state. To prepare an arbitrary superposition of the Floquet states $\alpha |\psi_+^{(N)}\rangle + \beta |\psi_-^{(N)}\rangle$, one must initialize the state in

$$|\psi(0)\rangle = \alpha |0\rangle + e^{i\phi t_f} \beta |2\rangle. \tag{S43}$$

Here $t_f$ is the preparation time and $\phi$ is a phase factor that accounts for the dynamical phase difference acquired by the two Floquet states throughout the adiabatic state preparation (see Section S4 C). Due to variations in the quasienergy spectrum throughout the state preparation, this dynamical phase is not easily obtained analytically. Instead, we obtain $\phi$ numerically for a fixed $N$ and $J/\omega_0$ driving scheme by performing reverse time-evolution, starting from the equal superposition state $(|\psi_+^{(N)}\rangle + |\psi_-^{(N)}\rangle)/\sqrt{2}$, and extracting the phase difference from the final superposition state $(|0\rangle + e^{i\phi t_f}|2\rangle)/\sqrt{2}$. By repeating this over a range of preparation times $t_f$, we find this phase factor to be $\phi \approx -3.08\omega_0 \times 10^{-2}$ for an $N=4$ and $J/\omega_0 = 2.5 \times 10^{-3}$ driving scheme.

Fig. S5 shows the results of preparing the approximate logical GKP states $|H_-\rangle_\mathcal{L}$, $|0\rangle_\mathcal{L}$ and $|1\rangle_\mathcal{L}$, where

$$|0\rangle_\mathcal{L} = \cos(\pi/8)|H_+\rangle_\mathcal{L} - \sin(\pi/8)|H_-\rangle_\mathcal{L}, \tag{S44a}$$
$$|1\rangle_\mathcal{L} = \sin(\pi/8)|H_+\rangle_\mathcal{L} + \cos(\pi/8)|H_-\rangle_\mathcal{L}, \tag{S44b}$$

are the logical computational basis states. Correspondingly, we choose $(\alpha, \beta) = (0, 1)$, $(\cos(\pi/8), -\sin(\pi/8))$ and $(\sin(\pi/8), \cos(\pi/8))$, respectively for the initial state in Eq. (S43). Similarly to Fig. 3 in the main text, we find that the squeezing and logical fidelity converge to those of the corresponding superposition of the Floquet states at longer preparation times (black dotted lines in Fig. S5), and are similarly affected by photon loss and flux noise to preparation of the $|H_+\rangle_\mathcal{L}$ state from the vacuum state. Whilst flux noise has a negligible impact on the squeezing of the prepared state in all cases, the logical fidelity of the prepared $|0\rangle_\mathcal{L}$ and $|1\rangle_\mathcal{L}$ states are slightly more sensitive to flux noise [Figs. S5(d) and S5(f)] than preparing the "natural" Hadamard states $|H_\pm\rangle_\mathcal{L}$ [c.f. Fig. 3(b) in the main text and Fig. S5(b)]. This is likely a result of flux noise leading to dephasing of the two Floquet states throughout the state preparation.

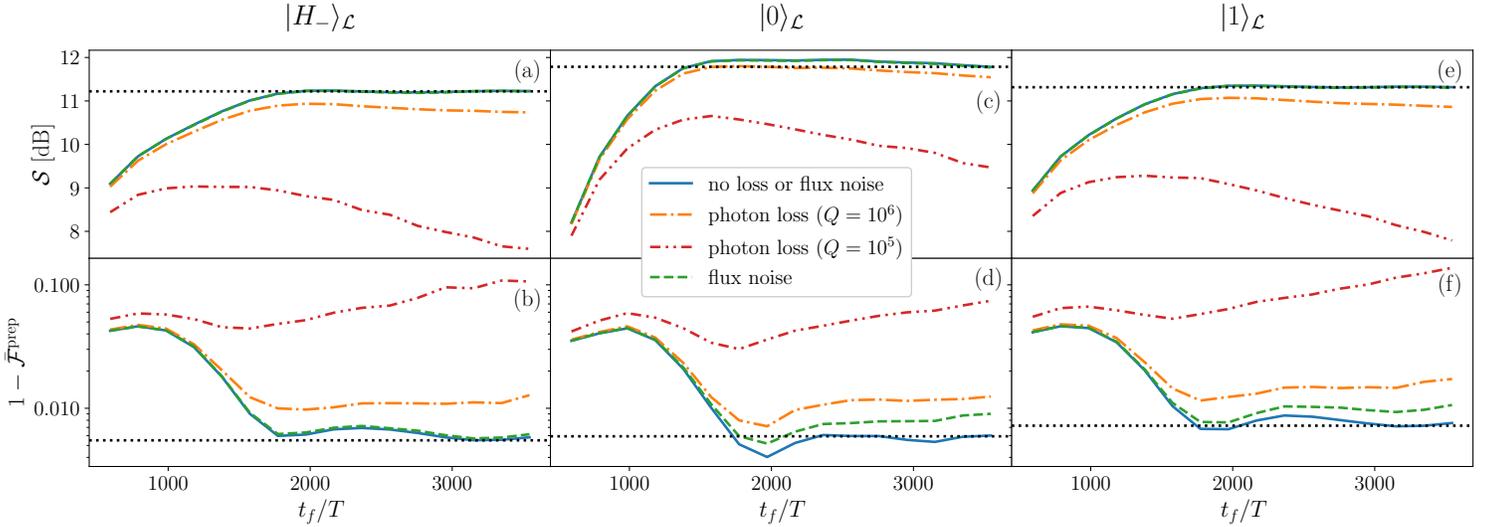

FIG. S5. Preparation of other GKP states. The first (second) row shows the squeezing (logical infidelity) of the prepared state, and the first, second and third columns correspond to an initial state given by Eq. (S43) with $(\alpha, \beta) = (0, 1)$, $(\cos(\pi/8), -\sin(\pi/8))$, and $(\sin(\pi/8), \cos(\pi/8))$, respectively. Logical infidelities are defined to be $\bar{\mathcal{F}}^{\text{prep}} = {}_\mathcal{L}\langle\Psi|\mathcal{D}[\hat{\rho}(t_f)]|\Psi\rangle_\mathcal{L}$, with $|\Psi\rangle_\mathcal{L} = |H_-\rangle_\mathcal{L}, |0\rangle_\mathcal{L}, |1\rangle_\mathcal{L}$ for the first, second and third columns, respectively. As in Fig. 3 of the main text, blue solid lines denote preparation with no photon loss and no flux noise, orange (red) (dot-)dot-dashed lines represent the preparation with photon loss at $Q = 10^6$ ($10^5$), and green dashed lines represent preparation with flux noise. The black dotted lines represent the values for the corresponding superposition of the Floquet states $\alpha|\psi_+^{(4)}\rangle + \beta|\psi_-^{(4)}\rangle$, with $(\alpha, \beta) = (0, 1)$, $(\cos(\pi/8), -\sin(\pi/8))$, and $(\sin(\pi/8), \cos(\pi/8))$ for the first, second and third columns, respectively. All results are for $J/\omega_0 = 2.5 \times 10^{-3}$ and $N = 4$.

## S6. FIDELITY OF PREPARED STATE WITH FLOQUET STATE

Whilst in the main text we have focused on the squeezing and logical fidelity as metrics to quantify how useful the prepared state is for GKP quantum computation, we may also compare how close the prepared state is to the corresponding Floquet state. For this, we use the following fidelity, $\mathcal{F}_+^{\text{prep}} = \langle \psi_+^{(4)} | \hat{\rho}(t_f) | \psi_+^{(4)} \rangle$, where $\hat{\rho}(t_f)$ is the prepared state. Note that this is agnostic to the GKP error-correction capabilities of the prepared state. With this metric, we observe the preparation infidelity to decrease monotonically in the lossless case (blue in Fig. S6), whilst the preparation infidelity in the presence of photon loss at $Q = 10^6$ ($10^5$) reaches a minimum at $t_f/T \approx 2000$ (1400) [orange (red) (dot-)-dot-dashed lines]. It is also interesting to note that with photon loss, the infidelities with the Floquet state are roughly an order of magnitude greater than the logical infidelities [c.f. Fig. 3(b) in the main text], whereas with flux noise, the infidelities with the Floquet state and logical infidelities are similar, albeit slightly higher. This demonstrates the GKP code's aptitude for correcting photon loss errors, since the decoder used to compute the logical fidelity corresponds to a round of ideal error correction [S5].

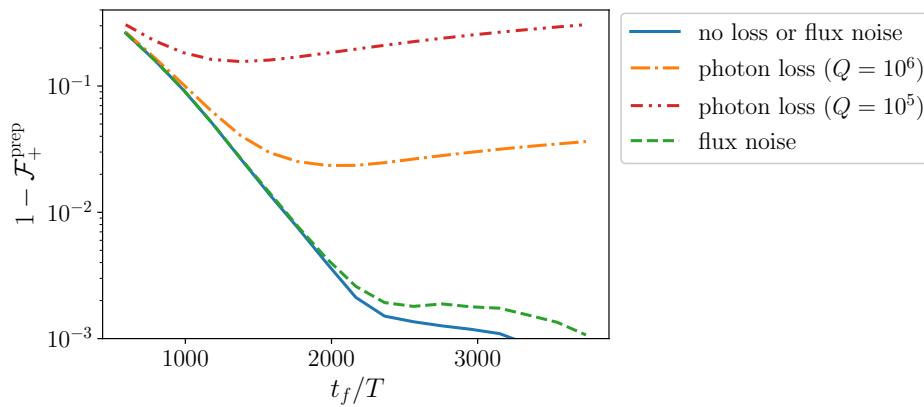

FIG. S6. Infidelity of the prepared state with the Floquet state $|\psi_+^{(4)}\rangle$ for different preparation times $t_f$. As in Fig. 3 in the main text, the blue solid line represents preparation with no photon loss and no flux noise, the orange (red) (dot-)dot-dashed line represents the preparation with photon loss at $Q = 10^6$ ($10^5$), and the green dashed line represents preparation with flux noise. Parameter values are $J/\omega_0 = 2.5 \times 10^{-3}$ and $N = 4$.

## S7. NUMERICAL SIMULATION OF PHOTON LOSS

Here we explain how we numerically simulate photon loss. Due to the high-energy nature of GKP states, the states considered in this paper require a large Fock truncation to simulate. Specifically, a Fock truncation of 200 was used. As such, simulating the full master equation [Eq. (7) in the main text] is computationally expensive. Rather than time-evolving $200 \times 200$ density matrices, we exploit the theory of quantum trajectories [S19] to integrate 200-dimensional state vectors, and use Monte Carlo sampling to construct the density matrix from the individual quantum trajectories.

Furthermore, for small $\kappa t_f$, the likelihood of a photon emission during the state preparation is small and may be calculated from the norm of the zero-photon trajectory by [S19]

$$p_{\text{loss}} = \langle \psi(0) | \hat{W}^\dagger(t_f) \hat{W}(t_f) | \psi(0) \rangle, \tag{S45}$$

where

$$\hat{W}(t) = \hat{\mathcal{T}} \exp\left[ -\frac{i}{\hbar} \int_0^t dt' \left( \hat{H}(t') - \frac{i\hbar}{2} \kappa \hat{a}^\dagger \hat{a} \right) \right] \tag{S46}$$

represents non-unitary evolution with photon loss. Specifically, we find the probability of at least one photon loss to occur during the state preparation to be less than 0.037 (0.31) at a quality factor of $Q = 10^6$ ($Q = 10^5$) (see Fig. S7). Therefore, to decrease simulation time further, we only simulate trajectories that have emitted at least one photon

and then weight these appropriately with the zero-photon trajectory:

$$\hat{\rho}(t_f) = \left|\bar{\psi}_0(t_f)\right\rangle\!\left\langle\bar{\psi}_0(t_f)\right| + \frac{1 - p_{\text{loss}}}{M} \sum_{i=1}^{M} |\psi_i(t_f)\rangle\langle\psi_i(t_f)|, \tag{S47}$$

where $\left|\bar{\psi}_0(t_f)\right\rangle = \hat{W}(t_f)\left|\psi(0)\right\rangle$ denotes the unnormalized zero-photon trajectory, and the $|\psi_i(t_f)\rangle$ are the (normalized) trajectories that have emitted at least one photon during the state preparation. To achieve a relative error of $\epsilon$, we need only simulate $M = p_{\text{loss}}(\sigma/\epsilon)^2$ trajectories, where $\sigma^2$ is the variance of the data. This is a factor of $p_{\text{loss}}$ less than if we were to employ standard Monte Carlo sampling algorithms [S20] that simulate many instances of the zero-photon trajectory. For all simulations we use a number of trajectories $M$ to achieve a relative error of $< 10^{-3}$.

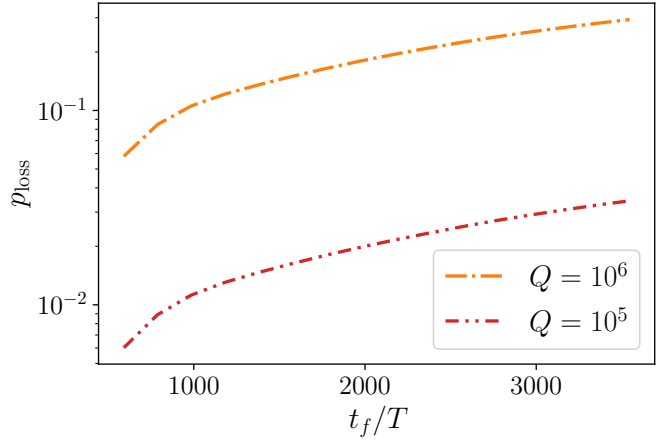

FIG. S7. Probability of at least one photon loss during state preparation as a function of preparation times at quality factors of $Q = 10^6$ (orange dot-dashed line) and $Q = 10^5$ (red dot-dot-dashed line). Parameters are the same as in Fig. 3 of the main text.

## S8. FLUX NOISE

In this section we provide additional details on flux noise. In Sections S8 A and S8 B we explain our model for simulating flux noise and the algorithm used to generate $1/f$ and white noise necessary for this model. In Section S8 C we show how the state preparation performs at different flux noise amplitudes, and in Section S8 D we compare simulating flux noise with the simplified and full circuit models.

### A. Flux noise model

We simulate the impact of a noisy flux by subjecting the driving function to $1/f$ (hereafter referred to as pink) and white noise via replacing $f(t)$ in Eq. (3) of the main text with

$$f(t) = 2 + \delta f_0(t) + \sum_{n=1}^{N}(4 + \delta f_n(t))\cos(4n\omega(t)t), \tag{S48}$$

which reduces to Eq. (4) of the main text with $\omega_0 = \omega(t)$ when $\delta f_0(t) = \delta f_n(t) = 0$. Here, $\delta f_0(t)$ represents noise in the DC component of the external flux, and $\delta f_n(t)$ represents noise in the AC components of the external flux [S21]. We model these noise components by sampling randomly generated numbers with a spectral density

$$S(\nu) = \frac{A_{\text{p}}^2}{|\nu|} + A_{\text{w}}^2, \tag{S49}$$

where the spectral density defines the two-time correlation function via $\langle \delta f_n(t_1)\delta f_n(t_2)\rangle = \int_{-\infty}^{\infty} d\nu\, S(\nu)e^{2\pi i\nu(t_1 - t_2)}$.

Experimentally measured values for flux noise are $A_\text{p}\epsilon/2\pi = 1.2 - 2.5 \times 10^{-6}$ [S22–S25] and $A_\text{w}\epsilon/2\pi = 3.1 \times 10^{-6}$ Hz$^{-1/2}$ [S25]. Thus, with $\epsilon = 1.25 \times 10^{-3}$, these values correspond to $A_\text{p} \approx 0.006 - 0.013$ and $A_\text{w} \approx 0.016$ Hz$^{-1/2}$. The flux noise results considered in the main text (green dashed lines in Fig. 3) corresponds to flux noise at the upper bounds of these estimates ($A_\text{p} = 0.013$ and $A_\text{w} = 0.016$ Hz$^{-1/2}$). We note that since the flux noise amplitudes scale with $1/\epsilon$, the sensitivity of the state preparation scheme to flux noise may be decreased further by increasing $\epsilon$. For a fixed ratio of $J/\omega_0$, this corresponds to decreasing $E_J/\hbar\omega_0$. Here and in the main text we choose $E_J/\hbar\omega_0 = 2$ so that a 1 GHz oscillator corresponds to a Josephson energy of $E_J/h = 2$ GHz. However, by using a higher frequency oscillator or a smaller Josephson energy, one may mitigate the effects of flux noise. In Section S8 C we investigate the state preparation quality at different flux noise amplitudes.

Due to the stochastic nature of the noise components $\delta f$, the squeezing and fidelity of all flux noise results have been averaged over 100 state preparations with the same noise amplitudes. We find this to result in a standard error ($\sigma/\sqrt{100}$, where $\sigma$ is the standard deviation) of less than one percent of the average.

### B. Pink and white noise generation

To generate random numbers with a spectral density according to Eq. (S49), we use the algorithm presented in Ref. [S26]. Here we explain this algorithm, and verify that it outputs the correct noise spectrum.

A real-valued noise process $\xi(t)$ with mean $\langle\xi(t)\rangle = 0$ and variance $\langle\xi(t)^2\rangle = \sigma^2$ may be described by its autocorrelation function [S27]

$$\langle\xi(t)\xi(t+\tau)\rangle = \frac{1}{t_f}\int_0^{t_f} dt\, \xi(t)\xi(t+\tau), \tag{S50}$$

where $t_f$ is the total time over which the noise process takes place. The spectral density may then be defined as the Fourier transform of the autocorrelation function

$$S(\nu) = \int_{-\infty}^{\infty} d\tau\, e^{-i2\pi\nu\tau} \langle\xi(t)\xi(t+\tau)\rangle. \tag{S51}$$

A white noise process is defined to be one whose spectral density is constant $S(\nu) = A_\text{w}$, whereas a pink ($1/f$) noise process is one whose spectral density scales inversely with the frequency $S(\nu) = A_\text{p}/|\nu|$. In order to simulate these noise processes, we must first discretize time into time steps $\Delta t$, yielding the discretized autocorrelation function and spectral density

$$\langle\xi[j]\xi[j+k]\rangle = \frac{1}{L}\sum_{j=0}^{L-1} \xi[j]\xi[j+k], \tag{S52a}$$

$$S[l] = \Delta t \sum_{k=0}^{L-1} \langle\xi[j]\xi[j+k]\rangle\, e^{-i2\pi kl\Delta t/L}, \tag{S52b}$$

where $\xi[j] = \xi(j\Delta t)$, $S[l] = S(l/L)$, $L = t_f/\Delta t$ is the number of time steps, and $j, k, l$ are integers.

To generate a discrete noise process with spectral density $S[l]$, we employ the algorithm in Ref. [S26]. This may be summarized as follows.

1. For each $j = 1, \ldots, \lfloor L/2 \rfloor$, define a complex variable $z[j] = z_1[j] + iz_2[j]$, where $z_1[j]$ and $z_2[j]$ are real-valued random variables drawn from a normal distribution with mean 0 and variance $S[j]/2$.

2. For each $j = 1, \ldots, \lfloor L/2 \rfloor$, define $z[-j] = z[j]^*$.

3. For $j = 0$, define $z[0]$ to be a random variable drawn from a normal distribution with mean 0 and variance $S[1]/2$.

4. Perform a discrete inverse Fourier transform on $z[j]$ to obtain the noise vector $\xi[j]$.

Note that $z[-j] = z[j]^*$ is necessary in order for the noise vector to be real-valued, and $S[1]/2$ is chosen to be the variance for the zero-frequency component in order to employ a zero-frequency cut-off, which is necessary for generating pink noise.

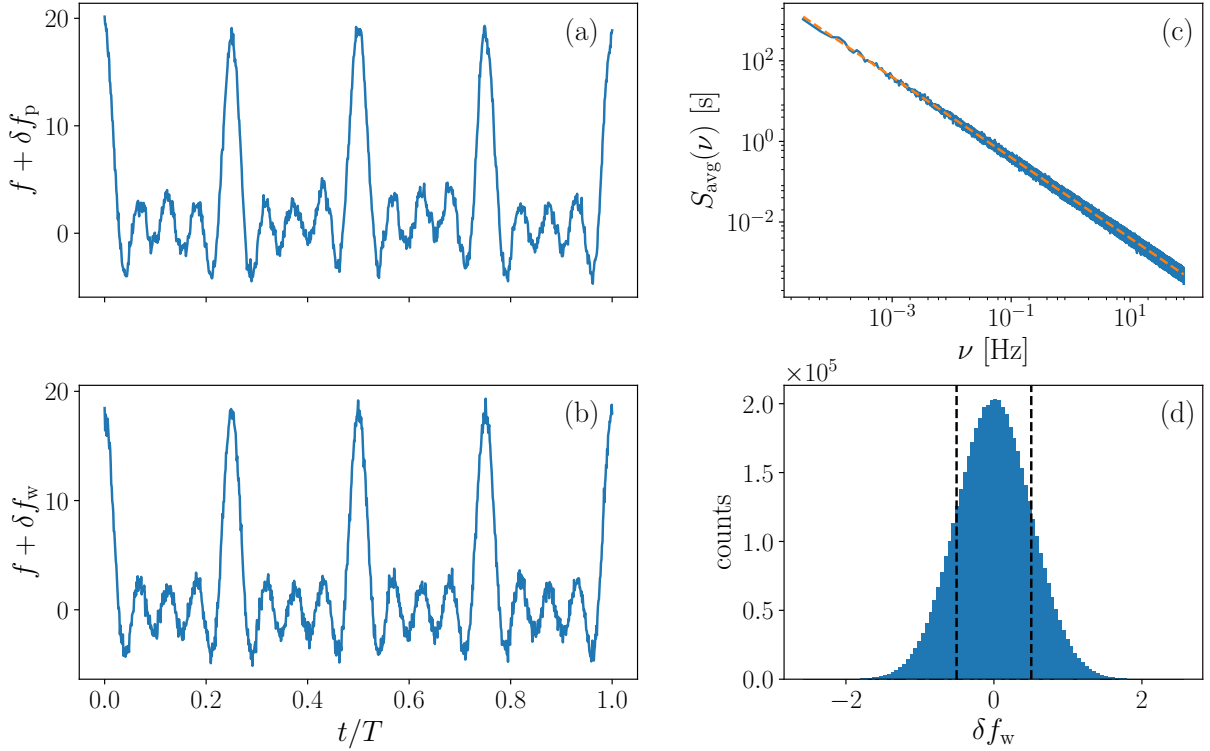

FIG. S8. Noise generation. (a) [(b)] Pink (white) noise vector $\delta f_\mathrm{p}$ ($\delta f_\mathrm{w}$) added to the ideal $N = 4$ driving function over a single HO period. $\delta f_\mathrm{p}$ ($\delta f_\mathrm{w}$) was generated using $A_\mathrm{p} = 0.2$, $A_\mathrm{w} = 0$ ($A_\mathrm{p} = 0$, $A_\mathrm{w} = 0.04$ Hz$^{-1/2}$), $\Delta t/T = 10^{-3}$, and $t_f/T = 5000$. (c) Average spectral density for 100 noise vectors generated using the same parameters as $\delta f_\mathrm{p}$. The orange dashed line is a linear fit to the average spectral density in log-log space with slope $m = -0.9999$ and intercept $b = -1.400$. (d) Histogram for the components of the noise vector $\delta f_\mathrm{w}$. The dashed black lines denote $\pm 1$ standard deviation $\sigma$, corresponding to a variance of $\sigma^2 = 0.255$.

For each noisy state preparation, $N+1$ vectors of length $L = t_f/\Delta t$ are generated using the above algorithm, and added to the ideal signal as per Eq. (S48). For our simulations, we set a time step of $\Delta t/T = 10^{-3}$ and set the maximum time step of the dynamics solver to be equal to $\Delta t$ so that every component of the noise vector is sampled.

As a demonstration, we compute pink and white noise vectors, $\delta f_\mathrm{p}$ and $\delta f_\mathrm{w}$, with a time step $\Delta t/T = 10^{-3}$ and length $L = 5 \times 10^6$, corresponding to $t_f/T = 5000$. For $\delta f_\mathrm{p}$ we use the noise amplitudes $A_\mathrm{p} = 0.2$, $A_\mathrm{w} = 0$, and for $\delta f_\mathrm{w}$ we use $A_\mathrm{p} = 0$, $A_\mathrm{w} = 0.04$ Hz$^{-1/2}$. Figs. S8(a) and S8(b) show the noise vectors after they have been added to the ideal signal for an $N = 4$ harmonic driving function, over a single HO period, where the first 1000 components have been sampled. That is, we set $\delta f_0 = \delta_\mathrm{p}$ or $\delta f_0 = \delta_\mathrm{w}$ and $\delta f_n = 0$ for $n \geq 1$ in Eq. (S48), which physically corresponds to only noise in the DC component of the signal.

To verify the algorithm generates pink noise correctly, the spectral density may be computed by taking the Fourier transform of $|\xi[k]|^2$. Fig. S8(c) shows the average spectral density for 100 pink noise vectors generated with the same parameters as $\delta f_\mathrm{p}$. The slope $m$ and intercept $b$ for a linear fit to the spectral density in log-log (base 10) space give the exponent and constant of proportionality $A_\mathrm{p}^2 = 10^b$, respectively, in $S(\nu) = A_\mathrm{p}^2 |\nu|^m$. We find values of $m = -0.9999$ and $10^b = 0.1995$ for this plot. These values converge to the analytical values of $m = -1$ and $A_\mathrm{p} = 0.2$ when the spectral density is averaged over more noise vectors, indicating that the algorithm outputs the correct noise spectrum.

To verify the algorithm generates white noise correctly, we plot a histogram of $\delta f_\mathrm{w}$ [Fig. S8(d)] and compute its variance to be $\sigma^2 = 0.255$. This value may be compared to its analytical value: note that from the inverse Fourier transform of Eq. (S52b), the variance of white noise ($S[l] = A_\mathrm{w}^2$) may be found by $\sigma^2 = \langle \xi[j]^2 \rangle = A_\mathrm{w}^2/\Delta t$. For $A_\mathrm{w} = 0.04$ Hz$^{-1/2}$ and $\Delta t/T = 10^{-3}$, this value is $\sigma^2 = 0.255$, which agrees with the numerically calculated value for the variance.

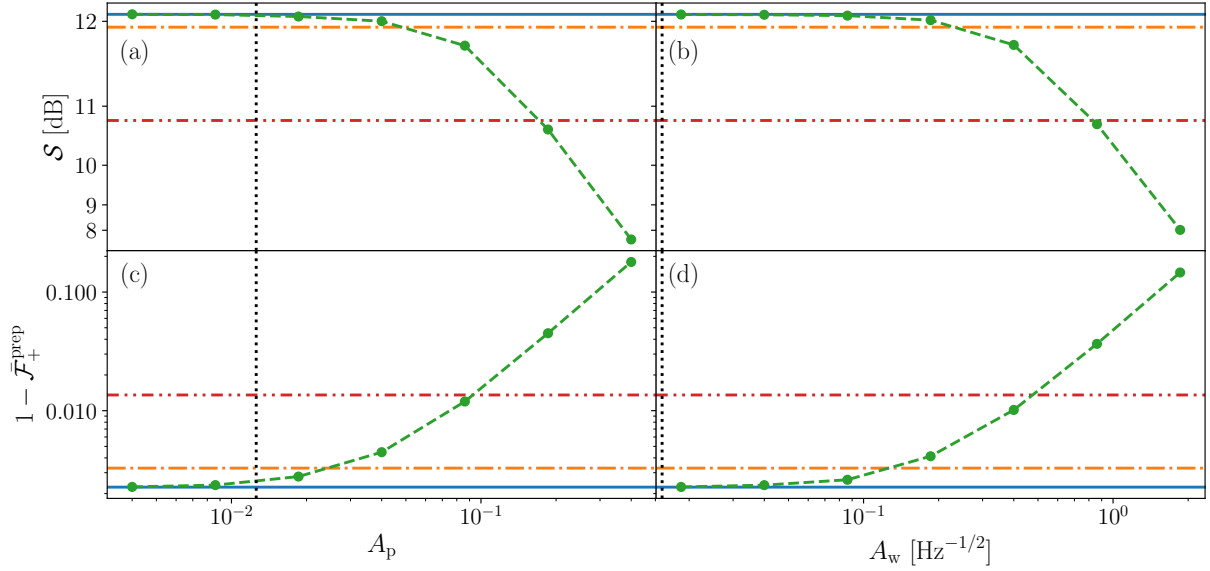

FIG. S9. Quality of prepared GKP state as a function of pink and white noise amplitudes at a preparation time of $t_f/T \approx 2000$. The first (second) row shows the squeezing (logical infidelity) of the prepared state and the first (second) column shows these as a function of pink (white) noise amplitudes. Green circles represent the values of squeezing and logical infidelity at different flux noise amplitudes; horizontal lines show the values with no loss (blue solid line), loss at $Q = 10^6$ (orange dot-dashed line), and loss at $Q = 10^5$ (red dot-dot-dashed line); and the vertical black dotted lines indicate the experimentally measured flux noise amplitudes used in Fig. 3. The other parameters are the same as in Fig. 3 of the main text ($J/\omega_0 = 2.5 \times 10^{-3}$ and $N = 4$).

### C. Different flux noise amplitudes

Here we analyze the quality of the prepared state as a function of the flux noise amplitudes. We consider pink and white noise separately at a fixed preparation time of $t_f/T \approx 2000$ and the same physical parameters as in Fig. 3 of the main text (green circles in Fig. S9). We find that it takes noise amplitudes greater than $A_p = 0.17$ or $A_w = 0.83$ Hz$^{-1/2}$ for the flux noise to impact both the squeezing and logical fidelity of the prepared state by the same amount as photon loss at $Q = 10^5$ (red dot-dot-dashed lines in Fig. S9). These values are more than an order of magnitude greater than the experimentally measured values for pink and white noise reported in Refs. [S22–S25] (black dotted lines in Fig. S9). Even with a quality factor of $Q = 10^6$, the experimentally measured pink (white) noise amplitudes are more than $4\times$ ($15\times$) less than those required for flux noise to impact both the squeezing and logical fidelity of the prepared state by the same amount as photon loss (orange dot-dashed lines in Fig. S9). Thus, photon loss is expected to be the dominant decoherence source for state preparation. The results also indicate that pink noise accounts for the majority of flux noise decoherence, and that white noise has little impact on the state preparation quality.

### D. Flux noise with the full circuit model

To decrease the complexity of our simulations, we have modeled flux noise using the ideal circuit model [Eq. (S26)] rather than the full circuit model [Eq. (S22)]. This is justified, because, under the same assumptions leading to Eq. (S26), as well assuming $C \ll C_J$ (which may be achieved physically by redistributing the shunt capacitance), the Hamiltonian with flux noise may be written as

$$\hat{H} = \hbar\omega_0 \hat{a}^\dagger \hat{a} - \frac{\hbar\omega_0}{\sqrt{\pi}}(\delta\varphi_{e1} + 2\delta\varphi_{e2})\hat{x} + E_J\left(\epsilon f(t) + \delta\varphi_{e1}\right)\cos\left(2\sqrt{\pi}\hat{x}\right), \qquad (S53)$$

where $\delta\varphi_{e1}$ and $\delta\varphi_{e2}$ represent the noise in each external flux. When $(\delta\varphi_{e1} + 2\delta\varphi_{e2})/\sqrt{\pi} \ll 1$, then the spurious linear $\hat{x}$ term may be neglected by a rotating-wave approximation [note that this does not apply to the non-linear $\cos(2\sqrt{\pi}\hat{x})$ term since $\cos(2\sqrt{\pi}\hat{x})$ also contains slow-oscillating terms in the rotating frame]. Since $\delta\varphi_e \sim \epsilon A$, where $A = A_w, A_p$

are the noise amplitudes [Eq. (S49)], the rotating-wave approximation is well-satisfied for the noise amplitude values considered here. Thus, Eq. (S53) reduces to Eq. (S26) with $\delta f$ in Eq. (S48) given by $\delta\varphi_{e1}/\epsilon$.

In Fig. S10 we verify that the numerical results are unaffected if we do not make the above simplifications and instead simulate state preparation using Eq. (S22) rather than Eq. (S26), with independently generated noise vectors for each external magnetic flux. We find that the quality of the prepared state is negligibly affected, and that the distribution of the capacitance between the SQUID and inductor [which does not appear in Eq. (S26)] has no impact on the results.

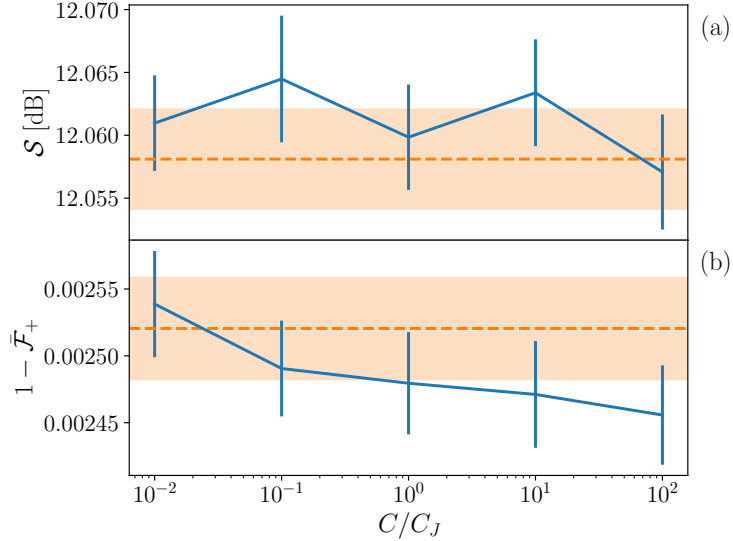

FIG. S10. Noisy state preparation simulated with different circuit models. (a) [(b)] Squeezing (logical infidelity) of the prepared state as a function of capacitance distribution $C/C_J$. The blue solid line denotes state preparation simulated using the full circuit model [Eq. (S22)], whereas the orange dashed line represents state preparation simulated using the simplified circuit model [Eq. (S26)]. The blue error bars and orange shaded region represent 1 standard error over 100 repetitions ($\sigma/\sqrt{100}$, where $\sigma$ is the standard deviation). Noise amplitudes are $A_p = 0.013$ and $A_w = 0.016$ Hz$^{-1/2}$ (same as for Fig. 3 in the main text and Figs. S5 and S6), a preparation time of $t_f/T = 2000$, and an $N = 4$, $J/\omega_0 = 2.5 \times 10^{-3}$ harmonic driving scheme.

## S9. LIFETIME OF THE GKP FLOQUET STATES

In this section we analyze the lifetime of a GKP Floquet state and show that it is decreased by the periodic drive. In the following we focus on the $|\psi_+^{(N)}\rangle$ state but all analytical arguments hold for the other GKP Floquet state $|\psi_-^{(N)}\rangle$ by swapping $+ \leftrightarrow -$. Here we set $\hbar = 1$.

First, we derive the decay rate for a GKP state initialized in one of the GKP Floquet states $|\psi_+^{(N)}\rangle$. Moving to a reference frame that is stationary with respect to the micromotion operator (defined in Section S1 A), the master equation [Eq. (7) in the main text] may be written as

$$\dot{\hat{\rho}}_K(t) = -i[\hat{H}_F, \hat{\rho}_K(t)] + \kappa\left(\hat{a}_K(t)\hat{\rho}_K(t)\hat{a}_K^\dagger(t) - \frac{1}{2}\{\hat{a}_K^\dagger(t)\hat{a}_K(t), \hat{\rho}_K(t)\}\right), \tag{S54}$$

where $\hat{O}_K(t) = e^{i\hat{K}(t)}\hat{O}(t)e^{-i\hat{K}(t)}$ for any operator $\hat{O}(t)$.

We now work in the $J/\omega_0 \to 0$ limit where $\hat{H}_F = \hat{H}_{\rm GKP}^{(N)}$ and the Floquet states $|\psi_i^{(N)}\rangle$ are exactly the eigenstates of $\hat{H}_{\rm GKP}^{(N)}$. Furthermore, for time scales much greater than a HO period $t \geq 1/J \gg 1/\omega_0$, we may neglect the dynamics due to the micromotion operator. This may be seen from Eqs. (S1a) and (S1b) since for $t \geq 1/J$ we have that $\hat{H}_F^{(1)}t = O(1)$, whereas $\hat{K}^{(1)}(t) = O(J/\omega_0)$ [S28] so that in the $J/\omega_0 \to 0$ limit, $\hat{K}^{(1)}(t) = 0$. Therefore, in this regime, Eq. (S54) may be rewritten as

$$\dot{\hat{\rho}}(t) = -i[\hat{H}_{\rm GKP}^{(N)}, \hat{\rho}(t)] + \kappa\left(\hat{a}\hat{\rho}(t)\hat{a}^\dagger - \frac{1}{2}\{\hat{a}^\dagger\hat{a}, \hat{\rho}(t)\}\right). \tag{S55}$$

To compute the decay rate, we take the overlap $\langle \psi_+^{(N)} | \cdot | \psi_+^{(N)} \rangle$ in Eq. (S55) and expand in the Floquet basis $|\psi_i^{(N)}\rangle$ (equivalently, the eigenbasis of $\hat{H}_{\text{GKP}}^{(N)}$). This yields

$$\dot{\rho}_{++} = -\kappa \langle \hat{a}^\dagger \hat{a} \rangle_+^{(N)} \rho_{++} - \frac{\kappa}{2} \sum_{i \neq \pm} \left( \langle \psi_+^{(N)} | \hat{a}^\dagger \hat{a} | \psi_i^{(N)} \rangle \rho_{i+} + \text{h.c.} \right) + \kappa \sum_{i,j \neq \pm} \langle \psi_+^{(N)} | \hat{a} | \psi_i^{(N)} \rangle \langle \psi_j^{(N)} | \hat{a}^\dagger | \psi_+^{(N)} \rangle \rho_{ij}, \quad \text{(S56)}$$

where $\rho_{ij} = \langle \psi_i^{(N)} | \hat{\rho} | \psi_j^{(N)} \rangle$ is the $(i,j)$-th element of the density matrix $\hat{\rho}$ in the eigenbasis of $\hat{H}_{\text{GKP}}^{(N)}$, and $\langle \hat{a}^\dagger \hat{a} \rangle_+^{(N)} = \langle \psi_+^{(N)} | \hat{a}^\dagger \hat{a} | \psi_+^{(N)} \rangle$ is the mean-photon number of the $|\psi_+^{(N)}\rangle$ GKP state. The reason the $|\psi_+^{(N)}\rangle$ GKP state is not included in the sums in Eq. (S56) is due to the fact that $\langle \psi_+^{(N)} | \hat{a}^\dagger \hat{a} | \psi_-^{(N)} \rangle = \langle \psi_+^{(N)} | \hat{a} | \psi_-^{(N)} \rangle = 0$, a consequence of these states respecting Fourier transform symmetry with different eigenvalues $e^{i\pi \hat{a}^\dagger \hat{a}/2} |\psi_\pm^{(N)}\rangle = \pm |\psi_\pm^{(N)}\rangle$, and $e^{i\pi \hat{a}^\dagger \hat{a}/2} \hat{a} e^{-i\pi \hat{a}^\dagger \hat{a}/2} = -i\hat{a}$.

For a state initialized in the $|\psi_+^{(N)}\rangle$ state, $\rho_{++} = 1$ and all other terms in Eq. (S56) are 0 initially. Thus, the matrix element initially decays as $e^{-\kappa \langle \hat{a}^\dagger \hat{a} \rangle_+^{(N)} t}$. Numerically integrating Eq. (S55) with $N = 4$ and a quality factor of $Q = 10^6$ ($10^5$), we find that the decay fidelity $\mathcal{F}_+^{\text{decay}} = \langle \psi_+^{(4)} | \hat{\rho}(t) | \psi_+^{(4)} \rangle = \rho_{++}$ agrees with the analytical estimate for times

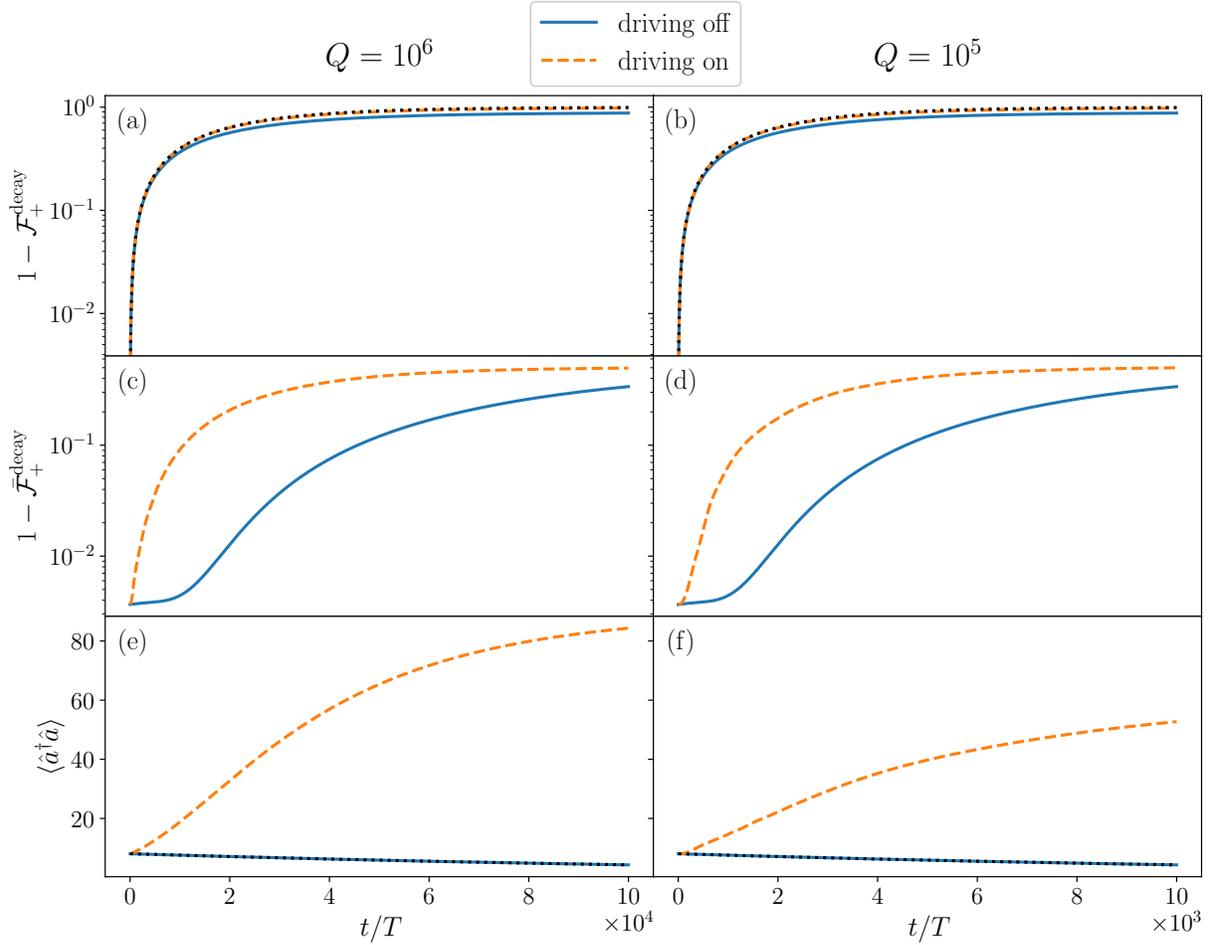

FIG. S11. Decay of the $|\psi_+^{(4)}\rangle$ GKP Floquet state with and without driving. (a,b) Infidelity between a time-evolved state $\hat{\rho}(t)$ and its initial state, where the state has been initialized in the $|\psi_+^{(4)}\rangle$ eigenstate of $\hat{H}_{\text{GKP}}^{(4)}$, and its time-evolution is defined by the master equation, Eq. (S55), with a quality factor of $Q = 10^6$ (left column) or $Q = 10^5$ (right column). The black dotted lines represent the analytical decay $e^{-\kappa \langle \hat{a}^\dagger \hat{a} \rangle_+^{(4)} t}$. (c,d) Logical infidelity of the time-evolved state $\hat{\rho}(t)$. (e,f) Mean photon number of the time-evolved state $\langle \hat{a}^\dagger \hat{a} \rangle = \text{Tr}[\hat{a}^\dagger \hat{a} \hat{\rho}(t)]$. The black dotted lines denote the analytical decay $\langle \hat{a}^\dagger \hat{a} \rangle_+^{(4)} e^{-\kappa t}$. Throughout, the blue solid (orange dashed) lines represent time evolution with the driving off (on), where the driving off is simulated by setting $\hat{H}_{\text{GKP}}^{(N)} = 0$ in Eq. (S55).

$0 \leq t/T \leq 10^5$ $(10^4)$ {orange dashed and black lines in Fig. S11(a) [Fig. S11(b)]}. Note that here we directly integrate the master equation rather than simulating trajectories since we are integrating a static Hamiltonian, which greatly reduces the complexity of the simulation.

When the driving is turned off, which may be simulated by integrating Eq. (S55) with $\hat{H}_{\text{GKP}}^{(N)} = 0$ since the effective Hamiltonian is 0, the decay fidelity improves slightly [Figs. S11(a) and S11(b)], and the logical fidelity $\{\bar{\mathcal{F}}_{\text{decay}} = \langle H_+| \mathcal{D}[\hat{\rho}(t)] |H_+\rangle\}$ improves substantially [Figs. S11(c) and S11(d)]. This is because, in the presence of photon loss, the driving significantly increases the mean photon number of the evolved state whereas with no driving the mean photon number decays as $\langle \hat{a}^\dagger \hat{a} \rangle_+^{(4)} e^{-\kappa t}$ [Figs. S11(e) and S11(f)]. This means that a photon-loss event is more likely when the driving is on, leading to a decrease in the decay fidelity.

Whilst only transitions to lower energy states are possible in a static system at zero temperature, transitions to any other Floquet state are possible for a driven system at zero temperature, since the drive can supply the energy for these transitions. This effects is known as Floquet heating, which is a well-studied phenomenon in the Floquet literature – see Ref. [S29] and the references therein, as well as Ref. [S30] where the same phenomenon is observed in a similar circuit to ours, and a method of combating this heating using colored dissipation [S31] is discussed.

## S10. QUASIENERGIES

Here we analyze the quasienergy characteristics of the harmonic driving scheme.

In Fig. S12(a) we plot the 3 quasienergies of Floquet states that are adiabatically linked to the 3 lowest HO eigenstates, along with 7 nearby quasienergies of Floquet states with a mean photon number < 60. We observe that the quasienergy gap between the GKP Floquet states (blue and orange circles) and the rest of the spectrum is small with respect to the energy scale defined by the HO frequency ($\Delta\varepsilon_{\text{gap}} = 7.82\hbar\omega_0 \times 10^{-4}$). This is due to the fact that the quasienergy spectrum for the bare HO is infinitely degenerate, and the non-linear perturbation to the HO spectrum is small ($J/\omega_0 \ll 1$). For $\kappa \geq 10^{-6}\omega_0$, this quasienergy gap offers no protection for the Floquet states, as observed in Fig. S11, where multiple transitions to Floquet states with higher mean photon number occur, resulting in the increase in mean photon number.

In Fig. S12(b) we plot this quasienergy gap $\Delta\varepsilon_{\text{gap}}$, as well as the quasienergy difference between the two Floquet states $\Delta\varepsilon_\pm = |\varepsilon_+ - \varepsilon_-|$ as a function of the number of harmonics $N$. We find that (with the exception of $N = 2$), the former quasienergy gap remains larger than the latter, and that both decrease with the number of harmonics. This is consistent with the fact that as $N$ increases, the effective Hamiltonian approaches the GKP Hamiltonian, whose spectrum is continuous.

The quasienergy spectrum also reveals that the Floquet state that the $|1\rangle$ state is adiabatically linked to corresponds

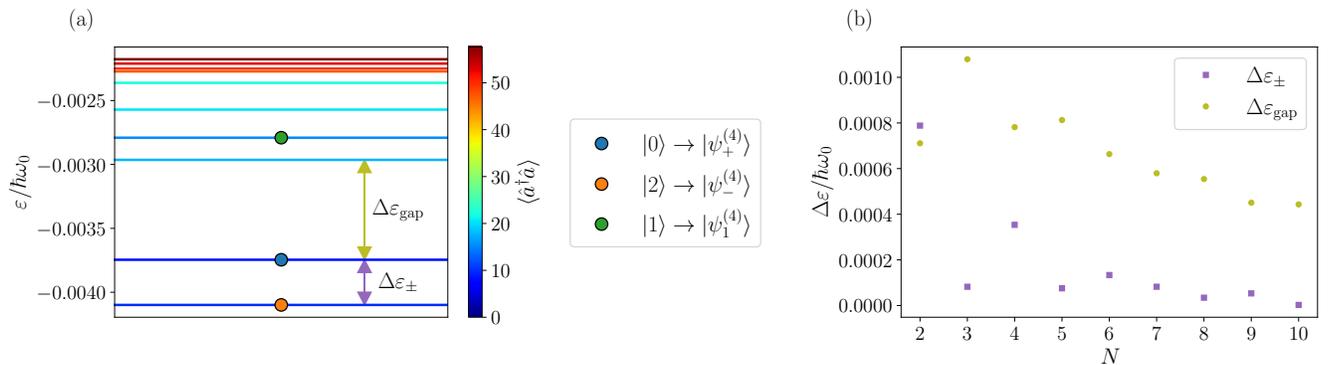

FIG. S12. Quasienergy characteristics of the harmonic driving scheme. (a) Quasienergy spectrum for the $N = 4$, $J/\omega_0 = 2.5 \times 10^{-3}$ harmonic driving scheme. Horizontal lines represent the 3 quasienergies of Floquet states that are adiabatically linked to the 3 lowest HO eigenstates, along with 7 nearby quasienergies of Floquet states with a mean photon number < 60, and have been color-coded according to the mean photon number of the respective Floquet state. The blue, orange and green dots denote the quasienergies for $|\psi_+^{(4)}\rangle$, $|\psi_-^{(4)}\rangle$ and the Floquet state that is adiabatically linked to the $|1\rangle$ HO eigenstate, respectively. (b) Quasienergy gaps as a function of $N$ for $J/\omega_0 = 2.5 \times 10^{-3}$ harmonic driving schemes. Purple squares denote the quasienergy gap between the two GKP Floquet states, and yellow circles denote the quasienergy gap between the quasienergy of $|\psi_+^{(4)}\rangle$ and the rest of the spectrum, shown by the purple and yellow arrows in panel (a).

to a quasienergy that is well-separated from those of the $|\psi_\pm^{(N)}\rangle$ states. This is also true of the $\{|0\rangle, |1\rangle, |2\rangle\}$ manifold at the beginning of the state preparation when $\omega(0) = \omega_0/(1 - \pi \times 10^{-2})$ so that $0 < 2\omega_0 \mod \omega(0) < \omega_0 \mod \omega(0)$. Thus, no quasienergy crossing between these states during the adiabatic state preparation is necessary. However, it is also possible for quasienergies to wrap around the Brillouin zone ($\varepsilon = \pm\pi\omega$) throughout the adiabatic process, leading to an array of other possible permutations of the quasienergy topology.

## S11. COMPARISON TO REF. [S1]

Here we provide a comparison of our state preparation scheme to the Floquet scheme for realizing GKP states introduced in Ref. [S1].

Ref. [S1] utilizes dynamical decoupling to design a periodic driving scheme that realizes GKP states and may be implemented in a superconducting circuit by shunting a Josephson junction with a high-impedance ($Z = 2R_Q$, as in our scheme) LC oscillator. The key difference in their physical realization is that the Josephson non-linearity is kept static, whilst the driving function is implemented through voltage pulses. On the one hand, this is experimentally advantageous because external voltages may be modulated with more control and less noise than external magnetic fluxes [S32]. On the other hand, there are a number of additional experimental difficulties with this proposal, which we elaborate on below.

Firstly, the driving function proposed in Ref. [S1] consists of a sequence of Dirac-delta functions, which require near instantaneous pulses to be applied to the voltage drive. Whilst no physical approximation to this driving function is provided in Ref. [S1], it may be possible to devise a harmonic approximation to the driving function in a similar vein to how the harmonic driving scheme in this letter approximates the kicked HO model. However, since each period of the driving function in Ref. [S1] consists of $\gtrsim 25$ unequally spaced voltage pulses to generate $\gtrsim 11$ dB GKP states, many more harmonics would be required to approximate this driving function than the, e.g., 4 harmonics used to approximate the 4 equally spaced Dirac delta functions in our scheme. This in turn increases the complexity of realizing the required driving scheme.

Secondly, the driving period is required to be very large with respect to the harmonic oscillator period. Whilst this is advantageous for approximating the Dirac-delta driving function using more harmonics (since more harmonics may fit into a single driving period given a fixed maximum physical driving frequency), it makes the scheme more sensitive to photon loss. Specifically, Ref. [S1] states that $T \geq \frac{2\pi}{\omega_0} 2^{4M}$, where $T$ is the driving period, $\omega_0$ is the harmonic oscillator frequency and $M$ is the number of "twirling rounds". To obtain comparably squeezed GKP states to our 4-harmonic driving scheme ($\mathcal{S} \approx 11$ dB), $M = 5$ twirling rounds are required, so that the driving period constraint requires $T \gtrsim \frac{2\pi}{\omega_0} \times 10^6$ (c.f. the requirement of $T = 2\pi/\omega_0$ for the harmonic driving scheme in this letter). This means $\kappa T$ will be increased by a factor of $10^6$ relative to the driving scheme proposed in this letter, at the same quality factor. Reference [S1] does not simulate state preparation of a GKP state, but it briefly outlines a possible ancilla-free method for state preparation utilizing their scheme. This involves replacing the Josephson junction with a $\cos(2\varphi)$ circuit element [S33] and evolving an initial state (e.g., the vacuum state) for multiple driving periods. Whilst no estimate of the number of driving periods required to prepare a GKP state is given, even a couple driving periods is $\sim 10^3$ times longer than our state preparation protocol, which has a preparation time of $t_f \sim 2000 \times 2\pi/\omega_0$ to prepare similarly squeezed GKP states.

Finally, the analysis performed in Ref. [S1] is done at the level of the rotating-wave approximation, which assumes that $E_J/\hbar\omega_0 \to 0$, where $E_J$ is the Josephson energy. Undoubtedly, a finite value of $E_J/\hbar\omega_0$ will lead to a degradation in the quality of the GKP states and it is unclear how low this value must be to obtain good-quality GKP states. Furthermore, since $E_J$ may not be made arbitrarily small, this may pose an experimental challenge. By contrast, our requirement of $J/\omega_0 \ll 1$ where $J = \epsilon E_J/\hbar$ may be achieved more easily by instead making the flux drive amplitude $\epsilon$ arbitrarily small whilst fixing $E_J$ to a more realistic value.

In view of the above, our driving scheme appears to be more favorable for state preparation as compared with that of Ref. [S1] due to the much faster preparation time. Nevertheless, adapting the state-preparation protocol developed here to the driving scheme of Ref [S1] could provide a very promising state-preparation scheme, due to its mere manipulation of a voltage drive. This would require nontrivial modifications to be made to the scheme of Ref. [S1] in order to achieve a more physical driving function and relax the constraints $T \geq \frac{2\pi}{\omega_0} 2^{4M}$ and $E_J/\hbar\omega_0 \to 0$. We leave this as an interesting direction for future work.

---

* xkol5366@uni.sydney.edu.au


\fi

\end{document}